\documentclass[12pt,a4paper]{article}
\usepackage[latin1]{inputenc}
\usepackage{amsmath}
\usepackage{amsmath} 
\usepackage{amsfonts}
\usepackage{amssymb}
\usepackage{float}
\usepackage{cite}
\usepackage{breqn}
\usepackage{graphics}

\newcommand{\plainfootnote}[1]{%
  \begingroup
  \renewcommand{\thefootnote}{}
  \footnotetext{#1}%
  \addtocounter{footnote}{-1}
  \endgroup
}

\usepackage{fontawesome}

\usepackage{booktabs}
\usepackage[table,xcdraw]{xcolor}
\usepackage{tikz}
\usepackage{adjustbox}

\DeclareGraphicsExtensions{.pdf,.png,.jpg}
\usepackage[colorlinks=true,linkcolor=blue,citecolor=red]{hyperref}

\makeatletter
\newcommand{\mathleft}{\@fleqntrue\@mathmargin0pt}
\newcommand{\mathcenter}{\@fleqnfalse}
\makeatother
\newcommand*{\SavedEqref}{}
\let\SavedEqref\eqref
\renewcommand*{\eqref}[1]{%
  \begingroup
    \hypersetup{
      linkcolor=linkequation,
      linkbordercolor=linkequation,
    }%
    \SavedEqref{#1}%
  \endgroup
}

\usepackage[a4paper, width = 160mm, top = 30mm, bottom = 25mm, 
bindingoffset = 0mm]{geometry}

\def\beq{\begin{equation}}
\def\eeq{\end{equation}}
\def\bea{\begin{eqnarray}}
\def\eea{\end{eqnarray}}

\begin{document}

 
\begin{center}
	{\large \bf Transmission through Cantor structured Dirac comb potential
}
\vspace{0.7cm}		
		
{\sf Mohammad Umar }
\plainfootnote{\faEnvelope\ aliphysics110@gmail.com, opz238433@opc.iitd.ac.in}

\bigskip

{\em
$^{}$Optics and Photonics Centre\\ Indian Institute of Technology Delhi\\ New Delhi 110016, INDIA\\}

\bigskip	


\vspace{1.0cm}	
\noindent {\bf Abstract}		
\end{center}

\noindent
In this study, we introduce the Cantor-structured Dirac comb potential, referred to as the Cantor Dirac comb (CDC-$\rho_{N}$) potential system, and investigate non-relativistic quantum tunneling through this novel potential configuration. This system is engineered by positioning delta potentials at the boundaries of each rectangular potential segment of Cantor potential. 
This study is the first to investigate quantum tunneling through a fractal geometric Dirac comb potential. This potential system exemplifies a particular instance of the super periodic potential (SPP), a broader class of potentials that generalize locally periodic potentials. Utilizing the theoretical framework of SPP, we derived a closed-form expression for the transmission probability for this potential architecture. We report various transmission characteristics, including the appearance of band-like features and the scaling behavior of the reflection coefficient with wave vector $k$, which is governed by a scaling function expressed as a finite product of the Laue function. A particularly striking feature of the system is the occurrence of sharp transmission resonances, which may prove useful in applications such as highly sharp transmission filters.

\medskip
\vspace{1in}
\newpage
	

\section{Introduction}
Tunneling in quantum physics, an elemental concept unveiled in 1928 \cite{nordheim, gurney},  has persistently engrossed the scientific fraternity. This phenomenon, indispensable for elucidating quantum behavior, has undergone substantial advancements owing to the concerted efforts of researchers who have significantly augmented our understanding of its foundational principles \cite{condon1931quantum, wigner1955lower, bohm2012quantum, albeverio2012solvable, razavy2013quantum, esaki, burstein, giaever, josephson, lauhon}. A salient focus within this domain is the investigation of one-dimensional scattering by finite periodic chains of non-overlapping barriers or wells. This inquiry underscores critical quantum mechanical phenomena such as tunneling and interference \cite{lee1989one, kalotas1991one}, which are paramount for deciphering the physics of lattices and superlattices in solid-state physics and electronic devices \cite{capasso1990resonant}, as well as optical multi-quantum well systems and X-ray reflection theories from amorphous superlattices \cite{rhan1993investigations}. The celebrated work of Kronig and Penney in 1930, elucidating the behavior of an electron in an infinite (or semi-infinite) one-dimensional periodic array of delta potentials \cite{kronig1931quantum}, continues to serve as an indispensable tool for explicating various complex physical properties of real materials. Reading and Siege in 1972 analyzed particle scattering from a finite chain of delta potentials of arbitrary position and strengths utilizing the momentum representation technique \cite{reading1972exact}. The scattering problem for a finite chain of equidistant symmetric potentials has been rigorously examined by numerous research groups throughout the twentieth century, employing a diverse array of methodologies and theoretical frameworks \cite{slater1953augmented, cvetic1981scattering}.\\
\indent
In 1992, Griffiths presented a closed-form expression for the transmission probability in the context of one-dimensional scattering from a series of delta potentials \cite{griffiths1992scattering} and showed the emergence of band structure even when the potential count is less. This work aligns with previous work of Kiang \cite{kiang1974multiple} on multiple scattering in a Dirac comb system. Moreover, Griffiths demonstrated a comprehensive and accessible exposition of wave theory in locally periodic media, highlighting its multifaceted applications across various branches of physics \cite{griffiths2001waves}. These applications include transverse waves on a weighted string, longitudinal waves on a loaded rod, electromagnetic waves in transmission lines, and plane electromagnetic waves propagating through stratified optical media. In 2018, the concept of super periodic potential (SPP) was introduced which is an explicit generalization of the locally periodic potential and author formulated the closed-form expression of the scattering coefficients (reflection and transmission) for this generalized potential along with the explicit construction of the transfer matrix of the super periodic system \cite{hasan2018super}. This paper investigates the transmission features of super periodic rectangular potentials and super periodic delta potentials. As an application of SPP he showed that Cantor and Smith-Volterra-Cantor (SVC) potential is a special case of SPP system. 
Recently, we have demonstrated that not only the Cantor potential but also the polyadic Cantor potential of minimum lacunarity constitutes a specific instance of SPP \cite{umar2024polyadic}. 
This demonstrates that the notion of super periodicity encompasses a broader and more generalized framework than traditional periodicity, serving as a superset within the family of symmetric potentials. Consequently, recognizing the family of Cantor potential as a super periodic potential enhances the theoretical understanding of tunneling amplitudes through such potentials, providing a comprehensive analytical formulation.\\
\indent
Cantor potential is a fractal potential, it is pertinent to discuss the fractals and self-similarity. Fractals are geometrical entities characterized by self-similarity and uniformity across their smaller components, a property that is fundamental to their structure and identity \cite{mandelbrot1982fractal, mandlebrot1984fractal,mandelbrot1987fractals,mandelbrot1998nature,takayasu1990fractals, pietronero2012fractals, barnsley1988fractals, falconer2004fractal, lofstedt2008fractal, barnsley2014fractals}. These entities are composed of smaller replicas of their overall form, resulting in a recursive and repetitive pattern. The concept of fractals was first introduced by mathematician Benoit B. Mandelbrot \cite{mandelbrot1982fractal}, originating from the recursive application of a mathematical operation on a geometric figure known as the initiator. Through repeated application of this operation, termed the generator, self-similar structures, or fractals, are formed. At each level of recursion, the resulting sub-components exhibit a complete resemblance to the original object, exemplifying the principle of self-similarity. This intrinsic self-similarity implies that fractals are scale-invariant, maintaining consistent properties across all scales in mathematical contexts. Furthermore, the relevance of fractals extends beyond mathematics into the natural world, where many naturally occurring objects can be effectively described using fractal geometry \cite{mandelbrot1982fractal, lofstedt2008fractal, lovejoy1985fractal, scholz1989fractals, thompson1991fractals, havlin1995fractals}. Readers who are interested in the concept of fractals are encouraged to explore the references provided in \cite{hurd1988resource}.\\
\indent
Cantor fractal potential is a foundational example in the fractal family. It starts with a rectangular potential barrier of height $V$ and length $L$ (stage $S=0$). At each subsequent stage ($S=1, 2, 3, \dots $), a fraction $\rho^{-1}$ of the length from the previous potential stage is removed from the center, where $\rho \in \mathbb{R}^{+}$ and $\rho>1$. The case of $\rho=3$ defines the standard Cantor fractal potential. Alternatively, the general Smith-Volterra-Cantor (GSVC or SVC-$\rho$) potential, where at each stage $S$, a fraction $\rho^{-S}$ of the length from the previous stage is removed from the middle of each segment, produces the SVC-4 potential when $\rho=4$. Unlike the Cantor potential, the SVC-$\rho$ system is not a fractal due to its lack of consistent self-similarity. The quantum mechanical scattering properties associated with the Cantor fractal potential have been comprehensively studied \cite{konotop1990wave, sun1991wave, bertolotti1994spectral, bertolotti1996transmission, lavrinenko2002propagation, chiadini2003self, hatano2005strong, sangawa2005resonance, esaki2009wave, guerin1996scattering, honda2006rigorous, chuprikov2006new, chuprikov2000transfer, sakaguchi2017scaling, ogawana2018transmission,monsoriu2005transfer, chuprikov2000electron, jaggard1990reflection, konotop1991transmission, takeda2004localization}. 
These studies primarily employ the transfer matrix method in quantum mechanics to determine the scattering coefficients and their related characteristics. Next, the polyadic Cantor system \cite{mandelbrot1998nature, falconer2004fractal, barnsley2014fractals} extends the classic Cantor set, providing a wider array of fractal configurations. Unlike the traditional Cantor set, which removes a single middle third segment at each iteration, the polyadic Cantor system removes multiple segments per iteration. This method produces fractal patterns with varying levels of complexity and lacunarity, a concept explored in depth by Falconer \cite{falconer2004fractal}. 
A fundamental aspect of the taxonomy of the polyaidc Cantor potential system is its association with a parameter called lacunarity \cite{mandelbrot1982fractal, mandelbrot1994fractal, mandelbrot1995measures, lin1986suggested, gefen1983geometric, allain1991characterizing, yasar2005fractal, jaggard1997polyadic, jaggard1998scattering, monsoriu2006quantum, villatoro2008tunneling}, which significantly impacts its structural characteristics.\\
\indent
As previously discussed, tunneling through periodic and super periodic delta potentials has been extensively studied. In this article, we aim to investigate a novel type of super periodic delta potential known as the super periodic Cantor-structured Dirac comb potential, herein referred to as the Cantor Dirac comb (CDC-$\rho_N$) potential system. To the best of our knowledge, the tunneling phenomena through this specific type of Dirac comb system have not been previously examined. The key parameter $N$ in the subscript of the term CDC-$\rho_{N}$ is associated with the number of Dirac potentials present at stage $S=2$, representing $2N$ delta potential counts at this stage. 
In the context of the polyadic Cantor potential family, the parameter $N$ denotes the potential count at the first stage $S=1$. The construction of the CDC-$\rho_{N}$ system involves positioning delta potentials at the boundaries of each Cantor potential segment at each stage. It is an extraordinary fact to see that the CDC-$\rho_{N}$ system constitutes a special case of SPP. The family of Cantor potentials commences at stage $S=0$, thereby exemplifying a specific instance of SPP. Correspondingly, the CDC-$\rho_{N}$ system initiates at stage $S=1$, ensuring its status as a specialized manifestation of SPP. Owing to differences in the inaugural stage, the polyadic Cantor potential system features $N^{S}$ potential segments at each stage $S$.  In contrast, each stage $S$ of the CDC-$\rho_{N}$ system encompasses $2N^{S-1}$ delta potentials.\\
\indent
We organize the paper as follows: Section \ref{section02} explores the concept of CDC-$\rho_{N}$ system. In Section \ref{section03}, we establish that the CDC-$\rho_{N}$ system constitutes a particular case of the SPP system. Section \ref{section04} presents the derivation of the transmission coefficient expression, while Section \ref{section05} offers an extensive graphical and in-depth analysis of the transmission characteristics. Finally, Section \ref{section08} concludes with a summary of the findings and outlines potential directions for future research.
\section{Cantor Dirac comb (CDC-\texorpdfstring{$\rho_{N}$}{N}) potential system}
\label{section02}
\begin{figure}[H]
\begin{center}
\includegraphics[scale=0.28]{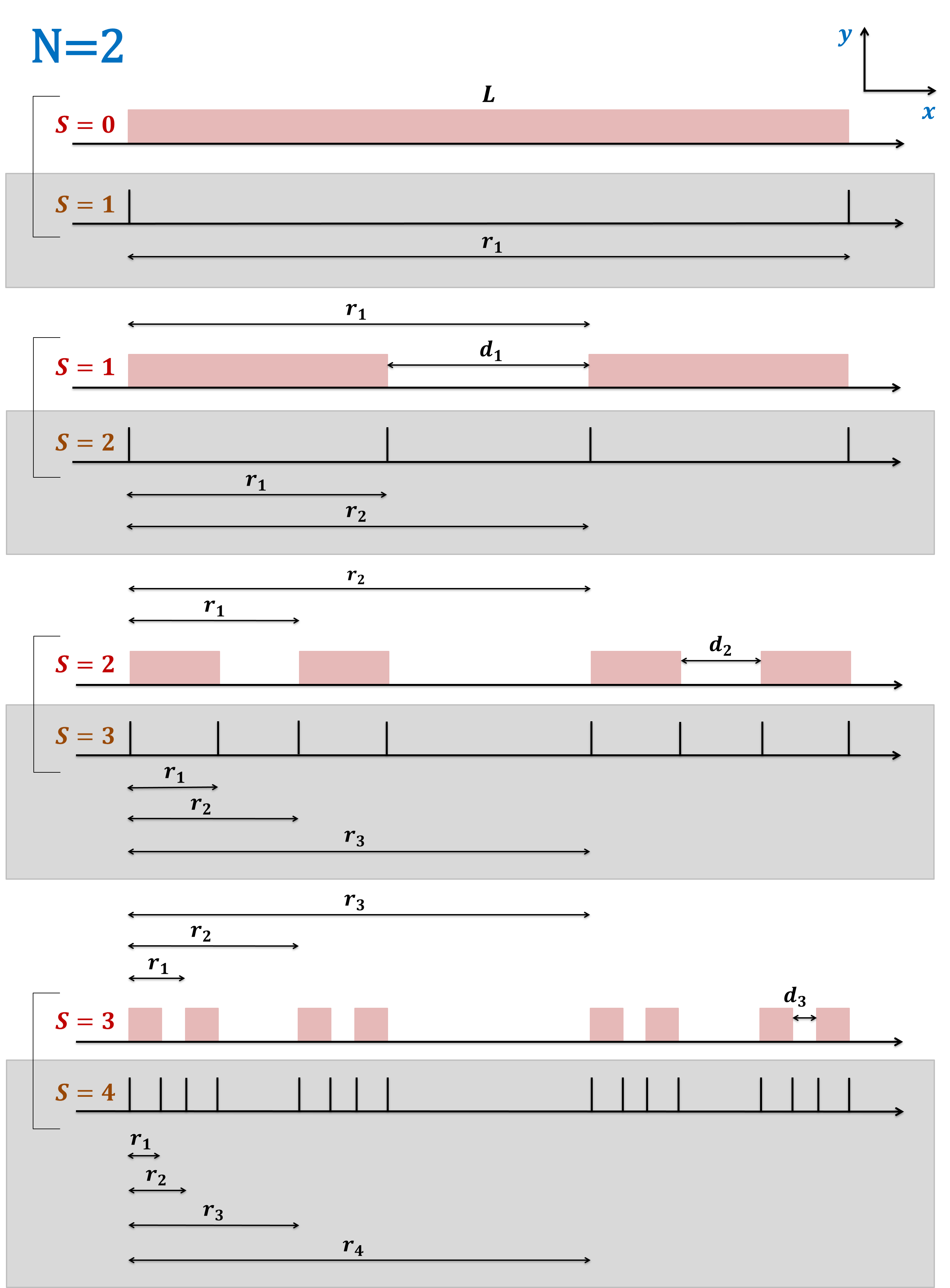} 
\caption{\it This figure illustrates the formation of the GC potential (depicted in a white background) and the CDC-$\rho_{2}$ system (depicted in a light gray background). It is important to note that stage \(S=0\) marks the initial stage of the GC potential, while stage \(S=1\) represents the initial stage of the CDC-$\rho_{2}$ system. The GC potential is constructed through an iterative process where a fraction $\rho^{-1}$ of the length of the potential segment at each stage is removed. The CDC-$\rho_{2}$ system is created by positioning delta potentials at the boundaries of each potential segment of the GC potential at every stage. Moreover, the construction of the CDC-$\rho_{2}$ system at each stage can also be seen as a super periodic repetition of delta potentials in a defined manner. The super-periodic distances, denoted by \(r_{q}\), are characterized by the relationship \(r_{q}=f(L, \rho)\).
}
\label{figure_01}
\end{center}
\end{figure}

\begin{figure}[H]
\begin{center}
\includegraphics[scale=0.465]{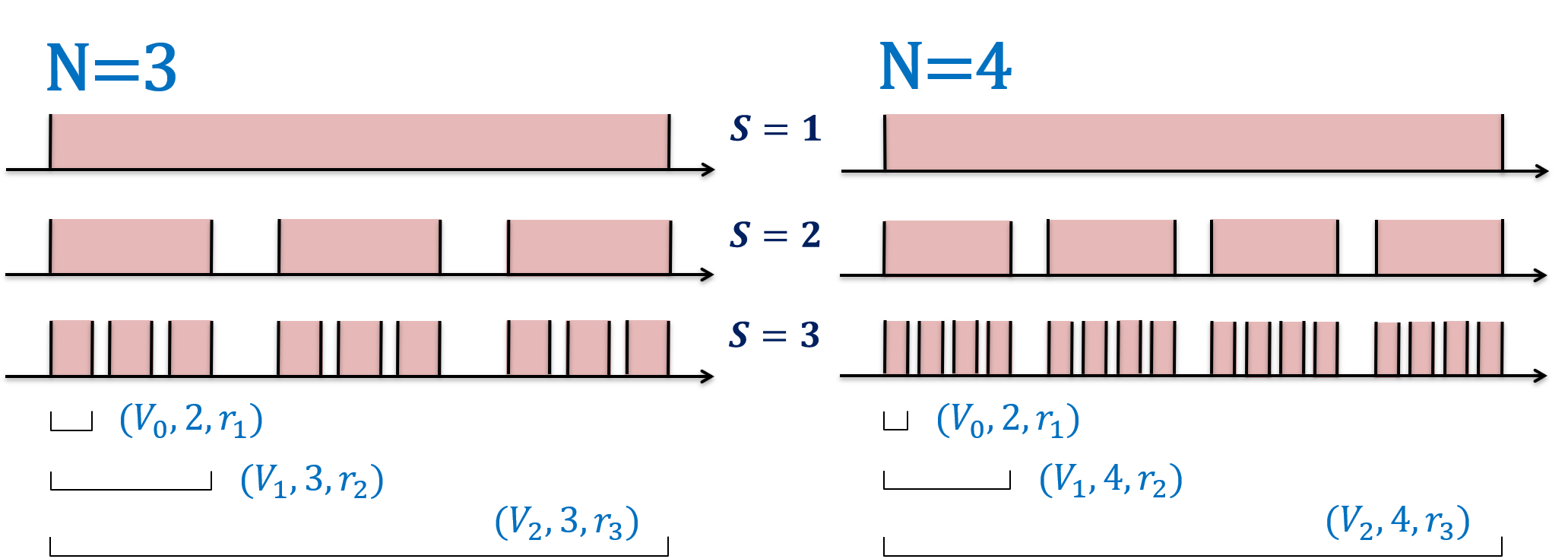} 
\caption{\it This figure shows the formation of CDC-$\rho_{3}$ and CDC-$\rho_{4}$ systems. The opaque regions represent the rectangular potential segment and the whole system represents the low lacunarity polyadic Cantor potential for \(N=3\) and \(N=4\), where \(N\) is the count of rectangular segments at the first stage (the second stage in the context of CDC-$\rho_{N}$ system. Delta potentials placed at segment boundaries create the polyadic Cantor-structured Dirac comb potential, as highlighted by the black vertical lines, thereby representing the CDC-\(\rho_{3}\) and CDC-\(\rho_{4}\) systems.}
\label{figure_02}
\end{center}
\end{figure}

To comprehend the concept of the CDC-$\rho_{N}$ potential system, it is essential to first explore the framework of the general Cantor (GC) potential system. The GC potential is constructed by modifying a rectangular barrier potential of finite length \( L \) and height \( V \equiv V_{0} \) through the iterative removal of a fraction \( \rho^{-1} \) of the potential length at each stage \( S \), where $\rho \in \mathbb{R}^{+}$ and $\rho>1$. The construction of the GC potential is depicted in Fig. \ref{figure_01} (geometry in the white background). It can be understood as follows: In the initial stage \( S=0 \), the potential extends the full length \( L \). At stage \( S=1 \), a fraction \( \rho^{-1} \) of the length \( L \) is removed from the center, resulting in two potential segments, each of length \( b_{1} \). At stage \( S=2 \), a fraction \( \rho^{-1} \) of the length \( b_{1} \) is removed from the center of each of the two segments, creating four segments of length \( b_{2} \). The process continues similarly at stage \( S=3 \), where a fraction \( \rho^{-1} \) of the length \( b_{2} \) is removed from the center of each segment, leading to eight potential segments of length \( b_{3} \). In general, at each stage \( S \), a fraction \( \rho^{-1} \) of the length of the potential segment at the previous stage is removed from the center of each segment, resulting in \( 2^{S} \) segments of equal length \( b_{S} \). The length of the segments at each stage is given by the expression
\begin{equation}
b_{S} = \frac{L\left(1-\rho^{-1}\right)^{S}}{2^{S}}.
\end{equation}
\indent
Next, the CDC-$\rho_{N}$ potential is a potential model constructed based on the model of the GC potential system. This system is formulated by placing the delta potentials at the boundaries of each potential segment of the GC system. A key distinction between the two systems is that the CDC-$\rho_{N}$ system initiates at stage $S=1$, unlike the GC system, which begins at stage $S=0$. This shift is necessary to preserve the super periodicity for the CDC-$\rho_{N}$ system. The construction of the CDC-$\rho_{N}$ system for $N=2$ is illustrated in Fig. \ref{figure_01} (geometry shown with a light gray background). At stage $S=1$, delta potentials are introduced at the boundary points of the GC potential of stage $S=0$, spanning the interval from $x=0$ to $x=L$, thus introducing two delta potentials at $x=0$ and $x=L$. At stage $S=2$, delta potentials are positioned at the boundaries of each segment of the GC potential corresponding to stage $S=1$, resulting in four delta potentials, separated by a central gap of length $L \times \rho^{-1}$. This process is recursively applied in subsequent stages, where at stage $S=3$, delta potentials are placed at the boundaries of each potential segment of the GC potential of stage $S=2$, leading to a total of eight delta potentials. By stage $S=4$, the system contains sixteen delta potentials. In general, the CDC-$\rho_{2}$ system for an arbitrary stage $S$ is constructed by placing delta potentials at the boundaries of each rectangular segment of the GC potential of stage $S-1$, resulting in a total of $2^{S}$ delta potentials at stage $S$.
\indent
Similarly, the polyadic Cantor-structured Dirac comb potential can be constructed. Fig. \ref{figure_02} illustrates the low lacunarity polyadic Cantor potential for \(N=3\) and \(N=4\) systems. In this context, \(N=3\) and \(N=4\) represent the number of rectangular potential segments at the first stage of the polyadic Cantor potential system. It is to be noted here that the stage numbering $S$ in Fig. \ref{figure_02} applies to the CDC-$\rho_{N}$ system, while the same stage in the Cantor potential system is represented by $S-1$. By positioning delta potentials at the boundaries of each potential segment of the polyadic Cantor system, the polyadic Cantor-structured Dirac comb potential system is formed. The black vertical lines in Fig. \ref{figure_02} indicate the locations of these delta potentials, thereby representing the CDC-\(\rho_{3}\) and CDC-\(\rho_{4}\) systems. For a given $N$, each stage of CDC-$\rho_{N}$ system contains $2N^{S-1}$ delta potentials. A detailed comparison of the parameters between the Cantor potential system and the CDC-$\rho_{N}$ system is presented in Table \ref{tab:comparison}.
\section{CDC-\texorpdfstring{$\rho_{N}$}{N} system: A special case of super periodic potential}
\label{section03}
In line with the SPP paradigm introduced in \cite{mh_spp}, we provide a concise explanation to ensure clarity in this paper. Starting with a \textit{unit cell} potential \(V_{0}\), a periodic structure is formed by repeating this \textit{unit cell} potential \(N_1\) times at intervals \(r_1\), resulting in the periodic potential \(V_1 = (V_{0}, N_1, r_1)\). This process is iterated by treating \(V_1\) as a new \textit{unit cell}, repeating it \(N_2\) times at intervals \(r_2\), forming \(V_2 = (V_1, N_2, r_2)\). Continuing this hierarchy, \(V_3 = (V_2, N_3, r_3)\) and \(V_4 = (V_3, N_4, r_4)\), and so forth, leading to the super periodic potential of order \(S\), denoted \(V_S = (V_{S-1}, N_S, r_S)\). This iterative construction produces increasingly complex SPP structures, generalizing the Cantor potential as a specific case. The super periodic formalism of Cantor potentials is explored in detail in \cite{hasan2018super, singh2023quantum, umar2023quantum, vsingh2023quantum, umar2024polyadic}.\\
\indent
In the previous section, we described the construction of the CDC-$\rho_{N}$ system by positioning the delta potentials at the boundaries of each rectangular potential segment of the GC and polyadic Cantor potential systems. It is also feasible to develop this potential system through super periodic formalism. Here, we outline the methodology for constructing the CDC-$\rho_{N}$ system at any stage in a super-periodic manner, demonstrating that the CDC-$\rho_{N}$ is a special case of super periodic potential. Now, refer to Fig. \ref{figure_02} only for the geometry of the CDC-$\rho_{2}$ system (light gray area). At stage $S=1$, a \textit{unit cell} delta potential, denoted by $V_{0}(x)$, is positioned at $x=0$ and periodically repeated with a count of $N_{1}=2$ at a distance $r_{1}=L$. At stage $S=2$, the potential $V_{0}(x)$ is once again located at $x=0$ and periodically repeated with $N_{1}=2$ at intervals of $r_{1}$, forming a system consisting of two delta potentials. This system, now defined as the \textit{unit cell} $V_{1}(x)$, is then periodically repeated at a distance $r_{2}$ with a count of $N_{2}=2$. At stage $S=3$, the \textit{unit cell} potential $V_{0}(x)$ is placed at $x=0$ and periodically repeated with $N_{1}=2$ at $r_{1}$. The resulting system, defined as \textit{unit cell} $V_{1}(x)$, is further periodically repeated at $r_{2}$ with $N_{2}=2$, producing the system $V_{2}(x)$, which is then repeated at $r_{3}$ with $N_{3}=2$. This recursive, super periodic arrangement describes the construction of the CDC-$\rho_{N}$ system at any arbitrary stage $S$. The super-periodic distance $r_{S}$ is a function of the finite extent $L$ and the scaling parameter $\rho$, i.e., $r_{S}=f(L, \rho)$, to maintain the locally finite extent of the super periodic Dirac comb at any stage $S$.\\
\indent
In summary, a CDC-$\rho_{2}$ system is constructed by iteratively applying the following operations: start with the potential $V_{0}(x)$, then obtain $V_{1}(x)=(V_{0}(x), N_{1}=2, r_{1}(L,\rho))$. Next, make $V_{2}(x)=(V_{1}(x), N_{2}=2, r_{2}(L,\rho))$, followed by $V_{3}(x)=(V_{2}(x), N_{3}=2, r_{3}(L,\rho))$. Continue this process up to stage $S$ to obtain $V_{S}(x)=(V_{S-1}(x), N_{S}=2, r_{S}(L,\rho))$. The resulting system $V_{S}$ represents the CDC-$\rho_{2}$ system. Similarly, CDC-$\rho_{3}$ $(N_{1}=2, N_{2, 3, 4,...., S}=3)$ and CDC-$\rho_{4}$ $(N_{1}=2, N_{2, 3, 4,...., S}=4)$ systems can be constructed as shown in Fig. \ref{figure_02}, with the initial periodic count always being $N_{1}=2$. Following the same criteria, we can construct the CDC-$\rho_{N}$ $(N_{1}=2, N_{2, 3, 4, \ldots, S}=N)$ system of any stage $S$. From the geometry of Fig. \ref{figure_01} and \ref{figure_02} it is easy to show that the first periodic distance at each stage is expressed through 
\begin{equation}
    r_{1}=\frac{L}{N^{S-1}}\left(1-\frac{N-1}{\rho}\right)^{S-1}
\end{equation}
and consecutive super periodic distances $r_{q}$, where $q\ge 2$ are expressed through
\begin{equation}
r_{q} = \frac{L\rho_{+}}{N^{S-q+1}}\left(1-\frac{N-1}{\rho}\right)^{S-q},
\end{equation}
where $\rho_{+}=\rho+\rho^{-1}$ and $N$ represents the count of the rectangular potential of a Cantor system at stage $S=1$. It is to be noted here that for a given $N$, there are $2N$ delta potentials at stage $S=2$ of the CDC-$\rho_{N}$ system. For $N=2$, the above expressions are simplified as
\begin{equation}
    r_{1}=L\left(\frac{\rho_{-}}{2}\right)^{S-1}
\end{equation}
and 
\begin{equation}
    r_{q}=\frac{L\rho_{+}}{2}\left(\frac{\rho_{-}}{2}\right)^{S-q},
\end{equation} 
where, $\rho_{-}=\rho-\rho^{-1})$. For the standard Cantor case $(\rho=3)$, above distances are expressed through
\begin{equation}
    r_{1}=\frac{L}{3^{S-1}}
\end{equation}
and
\begin{equation}
    r_{q}=\frac{2L}{3^{S-q+1}}=\frac{2r_{1}}{3^{2-q}}.
\end{equation}
A comparative analysis of parameters between the fractal Cantor potential system and the CDC-$\rho_{N}$ system is presented in Table \ref{tab:comparison}. The table provides details on the initial stage, number of potentials, potential widths, and super periodic distances, along with their corresponding equations.
\begin{table}[H]
\centering
\renewcommand{\arraystretch}{2}
\begin{tabular}{|c||c||c|}
    \hline
    \textbf{Parameter} & \textbf{Cantor system} & \textbf{CDC-$\rho_{N}$ system} \\ 
    \hline \hline \hline
    \textbf{Inaugural stage} & $S=0$ & $S=1$ \\ 
    \hline \hline
    \textbf{Potential count} & $N^{S}$ & $2N^{S-1}$ \\ 
    \hline \hline
    \textbf{Potential width} & $b_{S} = \frac{L}{N^{S}}\left(1 - \frac{N-1}{\rho}\right)^{S}$ & $0$ \\ 
    \hline \hline
    \textbf{Super periodic distance} & 
    \begin{tabular}{@{}l@{}}  
    $r_{q} = \frac{L\rho_{+}}{N^{S-q+1}}\left(1 - \frac{N-1}{\rho}\right)^{S-q}$ \\[1ex]
    \textit{where } $q=1,\ldots, S$
    \end{tabular} & 
    \begin{tabular}{@{}l@{}}  
    $r_{1} = \frac{L}{N^{S-1}}\left(1 - \frac{N-1}{\rho}\right)^{S-1}$ \\[1ex]
    $r_{q} = \frac{L\rho_{+}}{N^{S-q+1}}\left(1 - \frac{N-1}{\rho}\right)^{S-q}$ \\[1ex]
    \textit{where } $q=2,\ldots, S$
    \end{tabular} \\ 
    \hline
\end{tabular}
\caption{\textit{Comparison of parameters between the Cantor potential system and the CDC-$\rho_{N}$ potential system. The table includes details on the inaugural stage, potential count, potential width and super periodic distance with their respective equations.}}
\label{tab:comparison}
\end{table}

\section{Transmission probability from CDC-\texorpdfstring{$\rho_{N}$}{N} system}
\label{section04}
In this section, we will derive the expression for the transmission coefficient for the CDC-$\rho_{N}$ system. To begin with, we will first obtain the transmission coefficient for the \textit{unit cell} delta potential $V_{0}(x) =V\delta(x)$. The transfer matrix for this specific delta potential is given by \cite{griffiths2001waves}
\begin{equation}
    M(k)=\begin{pmatrix}
M_{11}(k) & M_{12}(k) \\
M_{21}(k) & M_{22}(k)  
\end{pmatrix}=\begin{pmatrix}
1+\xi & \xi \\
\xi^{*} & 1+\xi^{*}  
\end{pmatrix},
\end{equation}
where \(\xi = i\zeta\) and \(\xi^{*}\) is the complex conjugate of \(\xi\), with \(\zeta\) being expressed through $\zeta =\frac{mV}{\hbar^{2}k}$ and $k=\sqrt{2mE}$ with $m=1$ and $\hbar=1$. Therefore, for the delta potential, the magnitude and phase of the matrix element $M_{22}(k)$ is expressed through
\begin{equation}
    \vert M_{22}(k) \vert= \sqrt{1+\zeta^{2}}
\end{equation}
and 
\begin{equation}
    \tau =-\tan^{-1}\zeta
\end{equation}
respectively. The scattering from a periodic delta potential is covered in the literature \cite{griffiths1992scattering, griffiths2001waves}. Nonetheless, to ensure the completeness of this section, we present the expression for the transmission probability. The transmission probability through a finite periodic delta potential, also known as a Dirac comb, with $N_{1}$ periodic counts are 
\begin{equation}
    T(N_{1}, k)=\frac{1}{1+\left[\zeta U_{N_{1}-1}(\Gamma_{1})\right]^{2}},
    \label{eq5}
\end{equation}
where $U_{N}(\Gamma)$ is the Chebyshev polynomial of the second kind and $\Gamma$ is the argument of the Chebyshev polynomial and here $\Gamma_{1}$ is expressed through
\begin{equation}
    \Gamma_{1}=\cos k r_{1}+\zeta \sin k r_{1},
\end{equation}
where, $r_{1}$ represents the periodic distance. This completes the discussion about the periodic delta potential. In the previous section, we have shown that the CDC-$\rho_{N}$ system of any stage is a special case of an SPP system. Hence, now we will find the expression for the transmission probability for the CDC-$\rho_{N}$ system in the context of super periodicity. It is shown in the literature \cite{mh_spp} that if the transfer matrix $M(k)$ of a \textit{unit cell} potential is known then the transmission probability of SPP of order $S$ is expressed through
\begin{equation}
    T(N_{1}, N_{2}, N_{3},\ldots,N_{S}, k)=\frac{1}{1+\left[\vert M_{12}(k)\vert U_{N_{1}-1}(\Gamma_{1})U_{N_{2}-1}(\Gamma_{2})U_{N_{3}-1}(\Gamma_{3})\ldots U_{N_{S}-1}(\Gamma_{S})\right]^{2}}.
\label{eq14}
\end{equation}
Chebyshev polynomial of the second kind can also be expressed in terms of sinusoidal function as
\begin{equation}
    U_{N}(\Gamma)=\frac{\sin(N+1)\gamma}{\sin \gamma},
\end{equation}
where $\gamma$ is expressed through $\gamma=\cos^{-1}\Gamma$. Hence Eq. (\ref{eq14}) can be expressed as 
\begin{equation}
\begin{split}
    T(N_{1}, N_{2}, N_{3},\ldots,N_{S}, k) & = \frac{1}{1+\left[\vert M_{12}(k)\vert\frac{\sin N_{1}\gamma_{1}}{\sin \gamma_{1}}\frac{\sin N_{2}\gamma_{2}}{\sin \gamma_{2}}\frac{\sin N_{3}\gamma_{3}}{\sin \gamma_{3}}\ldots\frac{\sin N_{S}\gamma_{S}}{\sin \gamma_{S}}\right]^{2}} \\
    & = \frac{1}{1+\vert M_{12}(k)\vert^{2} \prod^{S}_{q=1}\frac{\sin^{2} N_{q}\gamma_{q}}{\sin^{2} \gamma_{q}}}\\
    & = \frac{1}{1+\vert M_{12}(k) \vert^{2} \prod^{S}_{q=1}L^{(S)}_{q}(\gamma_{q})}.
\end{split}
\label{eq10}
\end{equation}
In the above expression, $\gamma_{q}=\cos^{-1}\Gamma_{q}$ and $L(y)$ represents the Laue function expressed as $L(y)=\sin^{2}Ny/\sin^{2}y$ which is generally utilized in the X-ray diffraction analysis. Now, coming back to Eq. \ref{eq14}, various $\Gamma$s appearing in this equation are the argument of the Chebyshev polynomial which represents the Bloch phases of the corresponding fully developed periodic system. The expression for $\Gamma_{q}$ is expressed through \cite{mh_spp}
\begin{multline}
    \Gamma_{q}(k)=\vert M_{22}(k) \vert \cos \left[\tau - k \left \{ \sum_{p=1}^{q-1}(N_{p}-1)r_{p} - r_{q}\right \} \right ] {\prod_{p=1}^{q-1}U_{N_{p}-1}(\Gamma_{p})}\\+\sum_{h=1}^{q-1}\left [\cos \left\{k\left({\sum_{p=h}^{q-1}N_{p}r_{p}} - {\sum_{p=h+1}^{q}r_{p}}\right)\right\}U_{N_{h}-2}(\Gamma_{h}){\prod_{p=h+1}^{q-1}U_{N_{p}-1}(\Gamma_{p})}\right],
\label{eq17}
\end{multline}
where $\tau$ represents the argument of the matrix element $M_{22}(k)$ and $r_{p}$ is the super periodic distance which is discussed later in this section. The expression is also valid for $q$ = $1$ and $2$ provided we drop the terms when the running variable is more than the upper limit for the summation symbols and take the terms as unity when the running variable is more than the upper limit for the product symbols. Also, in the above expression $N_{0}=1$ and $r_{0}=0$. 
This concludes the discussion of the transmission probability through the CDC-$\rho_N$ system, wrapping up the formulation of the transmission coefficient.\\ 
\indent
The definition of the CDC-$\rho_N$ system says that the CDC-$\rho_2$ system has a super periodic count of two across all stages, i.e., $N_1 = N_2 = N_3 = \ldots = N_S = 2$. Given this condition ($N_p = 2$), Eq. (\ref{eq17}) simplifies as follows
\begin{equation}
    \Gamma_{q}(k) = 2^{q-1}\sqrt{1+\zeta^{2}}\cos\{\tau-k\chi_{1}(q)\}\prod^{q-1}_{p=1}\Gamma_{p}-\sum^{q-1}_{h=1}\left\{2^{q-h-1}\cos\{k\chi_{2}(q,h)\}\prod^{q-1}_{p=h+1}\Gamma_{p}\right\}, 
    \label{gamma_02}
\end{equation}
where $\chi_{1}(q)$ and $\chi_{2}(q,h)$ are given by 
\begin{equation}
    \chi_{1}(q)=\left(\sum^{q-1}_{p=1}r_{p}\right)-r_{q},
\end{equation}
\begin{equation}
    \chi_{2}(q, h)=\chi_{1}(q) - \chi_{1}(h)=\left(\sum^{q}_{p=h}r_{p}\right)-(2r_{q}-r_{h}).
\end{equation}
The following section is dedicated to exploring the transmission properties through the CDC-$\rho_N$ system, providing a comprehensive analysis of the transmission features and underlying concepts. 

\section{Transmission features}
\label{section05}
In Section \ref{section03}, we demonstrated that the CDC-$\rho_{N}$ system is a particular instance of the SPP system. In Section \ref{section04}, we derived the expression for the transmission coefficient for the CDC-$\rho_{N}$ system using the SPP formalism. To ensure comprehensive coverage in this draft, we will first elaborate on the overarching characteristics of the SPP system, followed by a discussion on the tunneling probability through the CDC-$\rho_{N}$ system.
\subsection{Transmission resonances in super periodicity}
\begin{figure}[H]
\begin{center}
\includegraphics[scale=0.460]{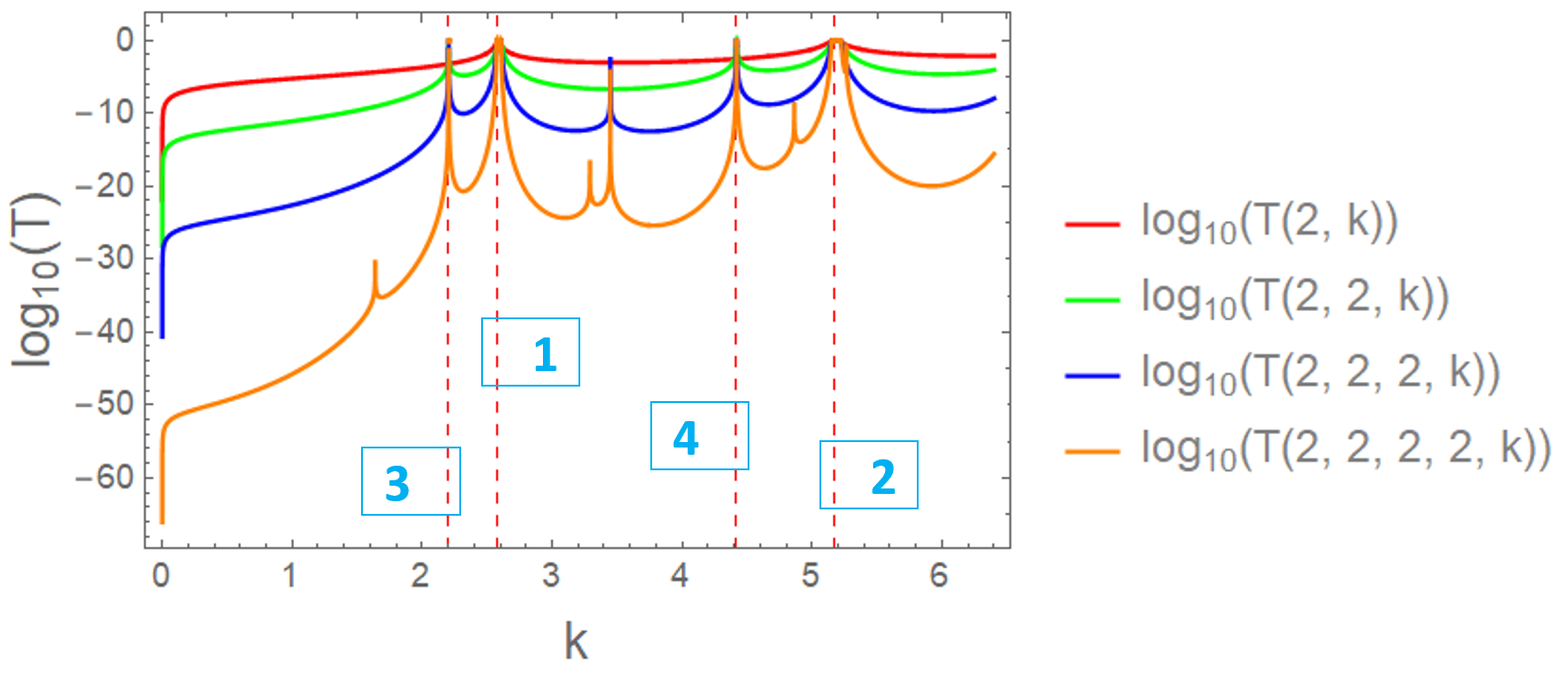}
\caption{\it Plots showing the transmission profile of the super periodic delta potential of order $S=1$, $2$, $3$ and $4$ with the super periodic distances $(r_{1}, r_{2}, r_{3}, r_{4}) = (1.15, 2.5, 4.5, 10)$. The super periodic count of delta potential for each order is mentioned in the plot legend of the figure. Here, the potential parameter is set to $V=15$. The grid lines labeled with the numbers $1$, $2$, $3$, and $4$ indicate the transmission resonance energy points \( k^{\star} \) with values $2.5835090$, $5.1747801$, $2.2032074$, and $4.4183635$ at the $k$-axis respectively. Transmission resonance points labeled as $1$ and $2$ arise due to the SPP of order $1$ and serve as transmission resonance points for subsequent SPPs. Similarly, resonance points labeled as $3$ and $4$ correspond to the SPP of order $2$, and they are resonance points for higher-order SPPs but not for the order $1$. A zoomed view of these transmission resonances is provided in Fig. $\ref{zoomresonance}$.}
\label{transmission_resonance_01}
\end{center}
\end{figure}

\begin{figure}[H]
\begin{center}
\includegraphics[scale=0.35]{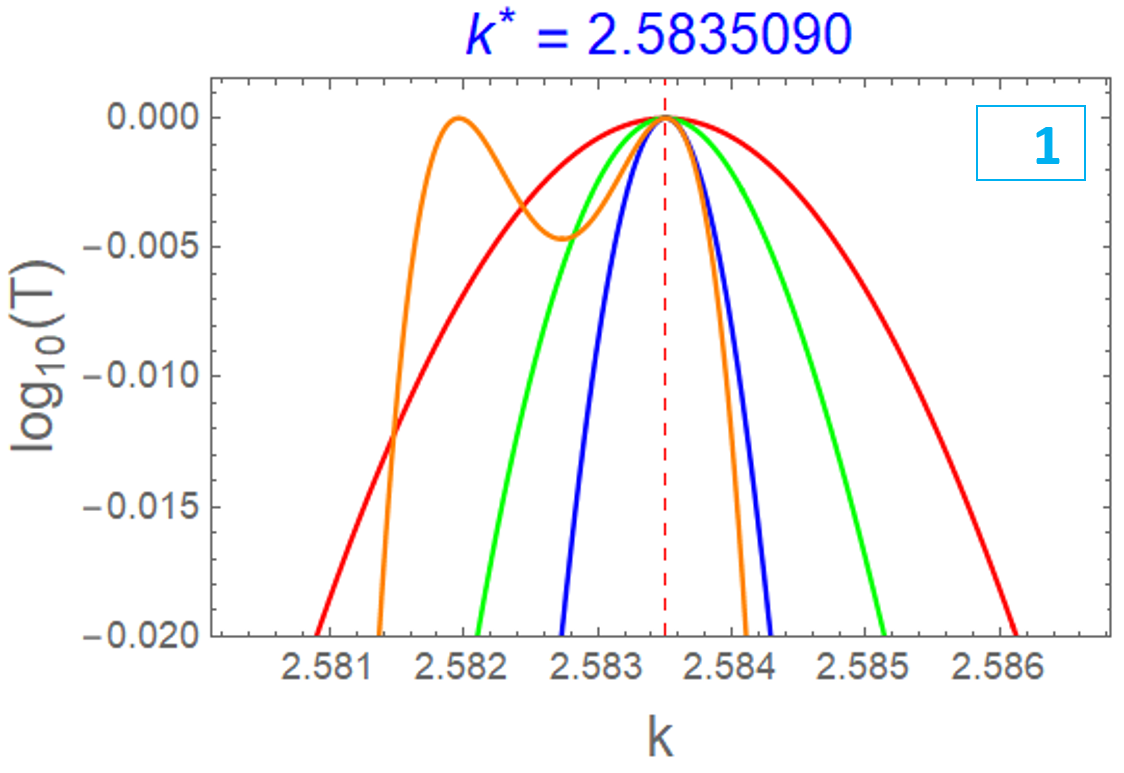} a
\includegraphics[scale=0.35]{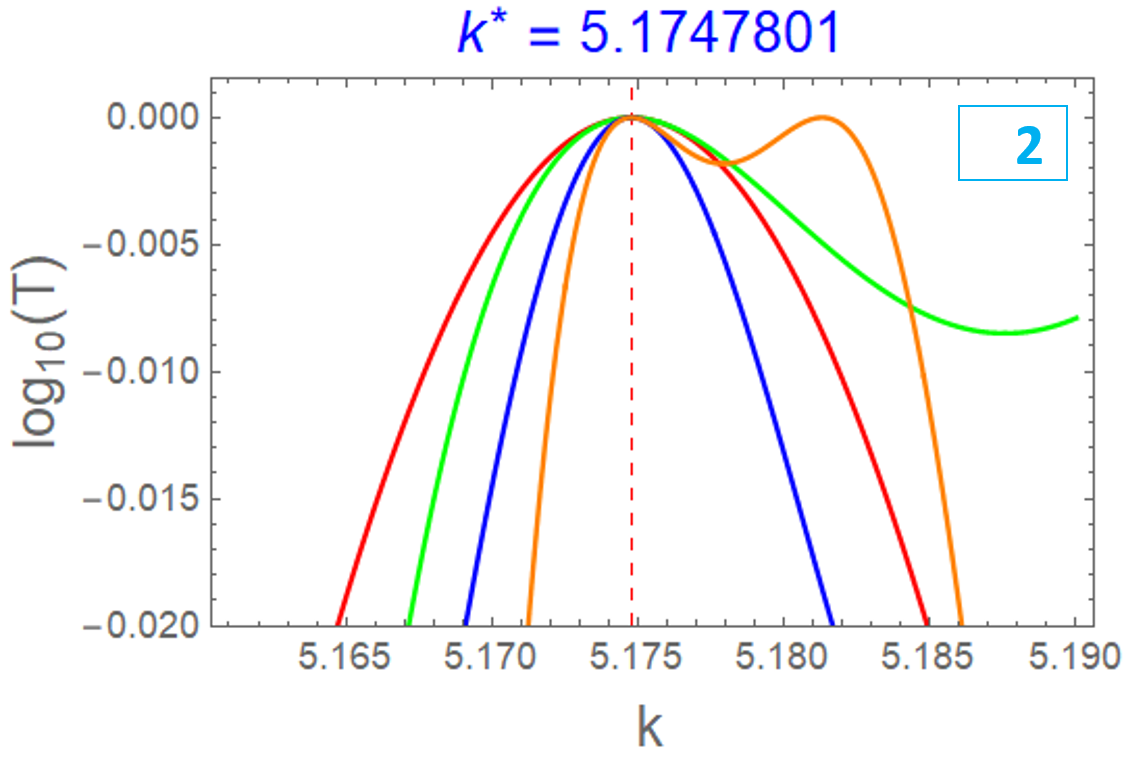}  b\\
\includegraphics[scale=0.35]{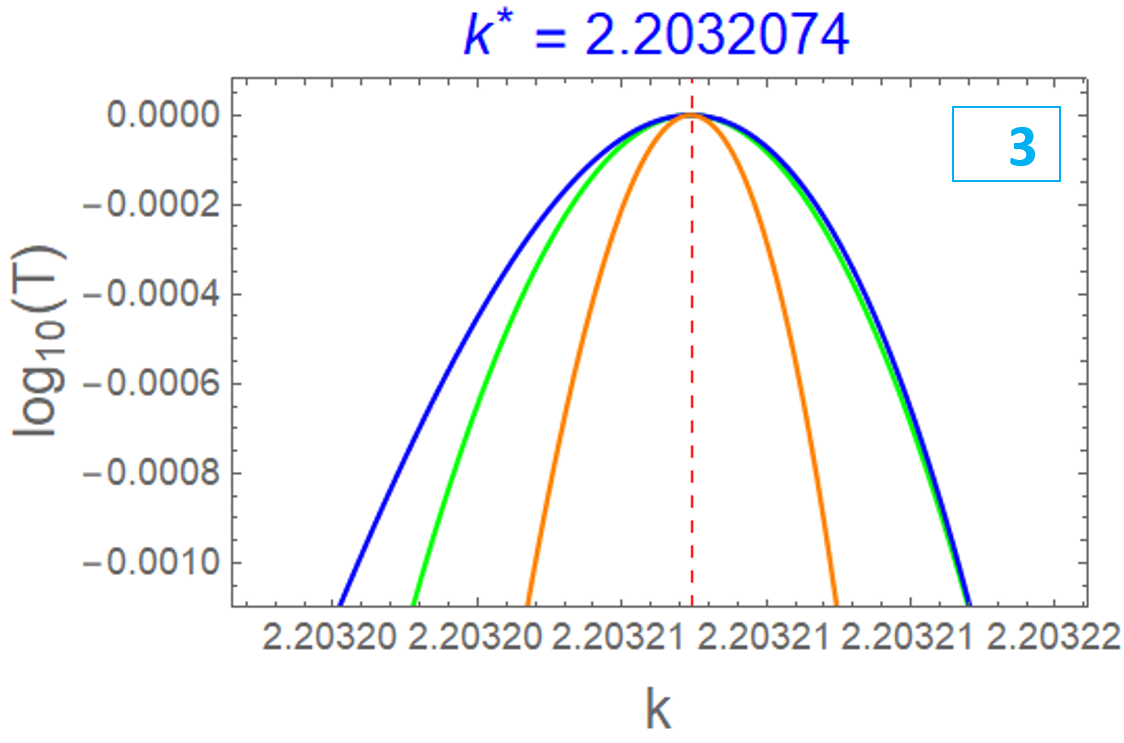} c
\includegraphics[scale=0.35]{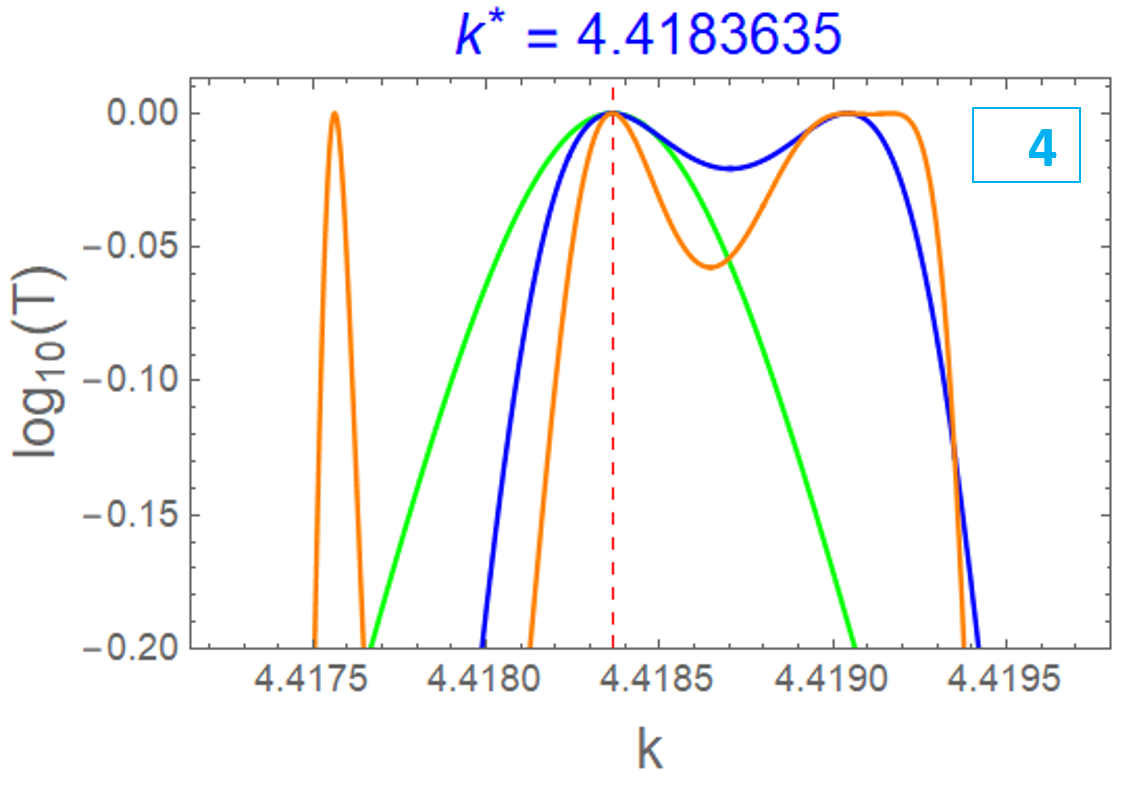} d
\caption{\it 
A magnified view of the transmission resonance points of Fig. $\ref{transmission_resonance_01}$, labelled as $1$, $2$, $3$, and $4$, is provided.}
\label{zoomresonance}
\end{center}
\end{figure}

Within the framework of the SPP system, the fundamental structure is defined by a \textit{unit cell} potential, whose characteristics are encapsulated within a $2\times 2$ transfer matrix. 
Let the zeros of the matrix element $M_{12}(E)$ be located at $E=E^\star$, i.e. $M_{12}(E^\star)=0$, where $E^\star$ signifies the transmission resonance energy. 
Then from the expression given in Eq. (\ref{eq14}), it becomes apparent that the energy $E^\star$ also serves as the transmission resonance energy for the SPP system of any order $S$. This can also be understood as follows. In the context of a \textit{unit cell} potential $V\equiv V_{0}$, if \( E^\star \) denotes the transmission resonance energy, it implies that a wave incident at this energy experiences perfect transmission. Extending this concept to a periodic system described by \( V_1 = (V_{0}, N_1, r_1) \), where \( N_1 \) represents the number of identical \textit{unit cells} and \( r_1 \) denotes the uniform spacing between these cells, \( E^\star \) also serves as the transmission resonance for the entire system. Thus, a wave with energy \( E^\star \) will propagate through the entire periodic arrangement with perfect transmission, as each \textit{unit cell} individually facilitates unimpeded passage at this resonance energy. Now if we consider this periodic arrangement $V_{1}$ as a \textit{unit cell} and periodically repeat this $N_{2}$ times at regular spacing interval $r_{2}$ to make $V_{2}=(V_{1}, N_{2}, r_{2})$. Then again the wave with energy $E^\star$ will pass through $V_{2}$ with the perfect transmission as this energy serves as the transmission resonance for the \textit{unit cell} $V_{1}$. With the same logic, the periodic potentials $V_{3}=(V_{2}, N_{3}, r_{3})$, $V_{4}=(V_{3}, N_{4}, r_{4})$, \ldots, $V_{S}=(V_{S-1}, N_{S}, r_{S})$ also allow to pass the wave with perfect transmission at energy $E^\star$ as this energy serves as the transmission resonance for the \textit{unit cells} $V_{3}$, $V_{4}$, \ldots ,$V_{S-1}$. The definition of SPP says that potential system $V_{S}$ is the SPP system of order $S$, hence it is conclusive that if energy $E^\star$ is the transmission resonance of a \textit{unit cell} potential then this energy unequivocally represents the transmission resonance for the SPP system of any order $S$, ensuring perfect transmission across the system.\\
\indent
Next, if for an energy $E=E_{n_{1}}$, $N_{1}\gamma_{1}=n_{1}\pi$, $n_{1}\in \mathbb{I}$, then $T(N_{1}, N_{2},\ldots,N_{S}, E_{n_{1}})=1$ for $S\ge 1$, $S\in \mathbb{I}^{+}$. Similarly, if for an energy $E=E_{n_{2}}$, $N_{2}\gamma_{2}=n_{2}\pi$, $n_{2}\in \mathbb{I}$, then $T(N_{1}, N_{2},\ldots,N_{S}, E_{n_{2}})=1$ for $S\ge 2$, $S\in \mathbb{I}^{+}$. Again, similarly , if for an energy $E=E_{n_{3}}$, $N_{3}\gamma_{3}=n_{3}\pi$, $n_{3}\in \mathbb{I}$, then $T(N_{1}, N_{2},\ldots,N_{S}, E_{n_{3}})=1$ for $S\ge 3$, $S\in \mathbb{I}^{+}$. Similarly, the same result applies at all those energies $E=E_{n_{S}}$ at which $N_{S}\gamma_{S}=n_{S}\pi$ implying that $T(N_{1}, N_{2},\ldots,N_{S}, E_{n_{S}})=1$. Each specified energy $E_{n_{S}}$ thus signifies a transmission resonance energy at which the quantum system exhibits optimal transmission, hence $E_{n_{S}}$ can be denoted as $E^\star_{n_{S}}$. In other words, this can be understood as follows: If SPP of order \( S = 1 \) with a periodic count \( N_1 \) exhibits certain transmission resonance energies, these energy points will also be transmission resonance points for subsequent SPPs of orders \( S = 2, 3, 4, \ldots, S \). Furthermore, if an SPP of order \( S = 2 \) with a periodic count \( N_2 \) exhibits transmission resonance energies, excluding those realized by the SPP of order \( S = 1 \), these energy points will likewise serve as transmission resonance points for higher-order SPPs of orders \( S = 3, 4, 5, \ldots, S \). Similarly, if an SPP of order \( S = 3 \) with a periodic count \( N_3 \) exhibits transmission resonance energies, excluding those due to the SPPs of orders \( S = 1 \) and $2$, these energy points will also be transmission resonance points for further SPPs of orders \( S = 4, 5, 6, \ldots, S \). This process continues iteratively up to the SPP of order \( S \).\\
\indent
The purpose of Fig. \ref{transmission_resonance_01} is to illustrate the transmission characteristics of the SPP system discussed in the preceding paragraph. Since the transmission coefficient $T$ ranges from $0$ to $1$ ($0 \le T \le 1$), the logarithmic function $f(T)=\log_{10}T$ can also be used to analyze the transmission features, ensuring $\log_{10}T \le 0$. In Fig. \ref{transmission_resonance_01}, we plot $\log_{10}T$ versus $k$ to display the transmission profiles of super periodic delta potential of orders $S = 1$, $2$, $3$, and $4$, corresponding to super periodic distances $(r_{1}, r_{2}, r_{3}, r_{4}) = (1.15, 2.5, 4.5, 10)$. The plot legend provides information about the super periodic count of the delta potential for each order, showing that each order has a super periodic potential count $N_{j} = 2$, $j$ ranging from $1$ to $S$. The potential parameter is set to $V = 15$ for this analysis. The plots feature grid lines labeled as $1$, $2$, $3$ and $4$, which denote the transmission resonance energy points $k^{\star}$ with values $2.5835090$, $5.1747801$, $2.2032074$, and $4.4183635$ respectively along the $k$-axis. Notably, resonance points labeled as $1$ and $2$ emerge due to the SPP of order $S=1$ and act as transmission resonance points for subsequent SPPs ($S>1$). Similarly, resonance points labeled as $3$ and $4$ arise from the SPP of order $S=2$ and serve as transmission resonance points for SPP of order $S=3$ and $4$ but not for SPP of order $S=1$. These features are prominently visible in the magnified view of the transmission resonances depicted in Fig. \ref{zoomresonance}. As shown in Figs. \ref{zoomresonance}c and \ref{zoomresonance}d, the absence of transmission 
\noindent
resonance for the SPP of order $S=1$ (represented by the red curve) is evident. This is because the transmission resonance points at $k^{\star}=2.2032074, 4.4183635$ correspond to the SPP of order $S=2$.

\subsection{Trasmission resonance band}
\label{section72}
When the argument \(\Gamma\) of the Chebyshev polynomial \(U_N(\Gamma)\) falls outside the interval \([-1, 1]\), divergence occurs as \(N \to \infty\). The arguments \(\Gamma_S\) (for \(S \in \mathbb{I}^+\)) used in these polynomials are inherently oscillatory due to the presence of trigonometric functions and terms derived from preceding arguments \(\Gamma_1, \Gamma_2, \Gamma_3, \ldots, \Gamma_{S-1}\) in the formulation of \(\Gamma_S\). Consequently, specific energy ranges exist where \(\Gamma_S\) exceeds the interval \([-1, 1]\). Within these energy ranges, the super periodic potential becomes opaque, resulting in zero transmission. These
\begin{figure}[H]
\begin{center}
\includegraphics[scale=0.39]{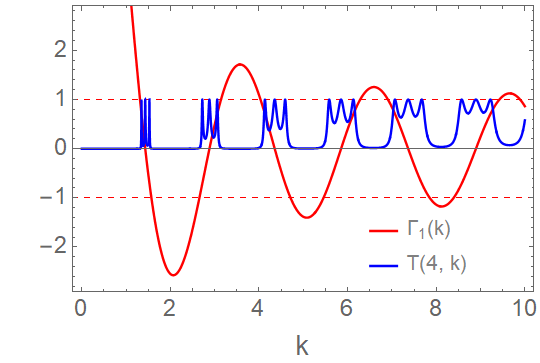} a
\includegraphics[scale=0.39]{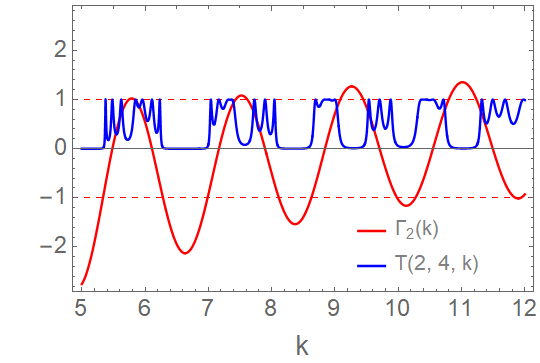}  b\\
\includegraphics[scale=0.39]{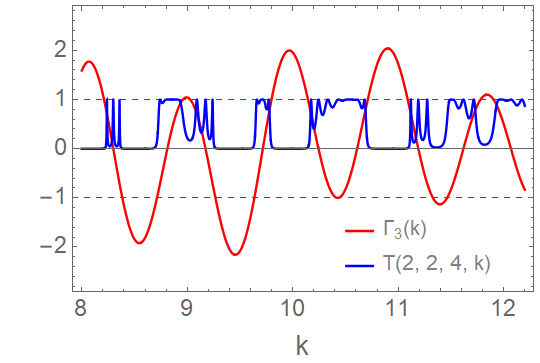} c
\includegraphics[scale=0.39]{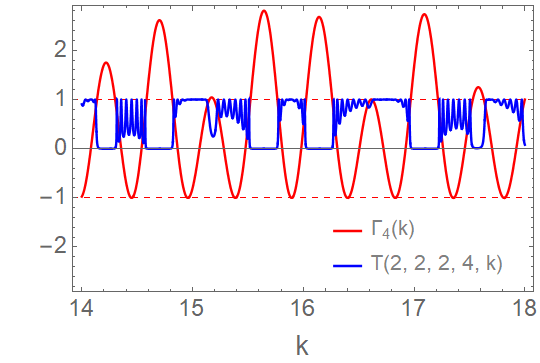} d
\caption{\it The transmission profile (blue curve) of the super periodic delta potential and the argument of the Chebyshev polynomial $\Gamma_{S}$ (red curve) are shown. The potential parameter is set at $V=5$, with super periodic distances of \(r_{1}=2\), \(r_{2}=3.5\), \(r_{3}=6.5\), and \(r_{4}=13\). In (a), we have the super periodic delta potential of order $S=1$ characterized by $(N_{1}, r_{1})=(4,2)$. In (b), the order is $S=2$ with $(N_{1}, N_{2}, r_{1}, r_{2})=(2,4,2,3.5)$. In (c), the order is $S=3$, defined by $(N_{1}, N_{2}, N_{3}, r_{1}, r_{2}, r_{3})=(2,2,4,2,3.5,6.5)$. Lastly, in (d), the order is $S=4$ with $(N_{1}, N_{2}, N_{3}, N_{4}, r_{1}, r_{2}, r_{3}, r_{4})=(2,2,2,4,2,3.5,6.5,13)$. Across all these cases, it is clear that the transmission probability nearly vanishes when $\vert\Gamma_{1}\vert>1$, $\vert\Gamma_{2}\vert>1$, $\vert\Gamma_{3}\vert>1$, and $\vert\Gamma_{4}\vert>1$.
}
\label{figure05}
\end{center}
\end{figure}
\noindent
ranges thus constitute forbidden bands for a particle confined within the super periodic potential. Therefore, the permissible energy spectrum of the particle exhibits a band structure, akin to the Kronig-Penney model. However, in this scenario, the band structure is uniquely modulated by the super periodicity of the potential.\\
\indent
The modulation of band structures, encompassing both permissible and forbidden energy bands, is graphically illustrated in Fig. \ref{figure05}. This figure shows the transmission probability of a super periodic delta potential up to order \(S=4\) (Fig. \ref{figure05}a-\ref{figure05}d), with a potential parameter of \(V=5\) and super periodic distances of \(r_{1}=2\), \(r_{2}=3.5\), \(r_{3}=6.5\), and \(r_{4}=13\). It is evident from these plots that the transmission probability nearly vanishes for a super periodic potential of order \(S\) when \(|\Gamma_{S}| > 1\). Fig. \ref{figure05}a represents a periodic delta potential with a periodic count \(N_{1}=4\) at a periodic distance \(r_{1}=2\), where the transmission probability nearly vanishes for \(\vert\Gamma_{1}\vert >1\). Similarly, Fig. \ref{figure05}b shows the transmission probability through a periodic Dirac comb with parameters \((N_{1}, r_{1}) = (2, 2)\) and \((N_{2}, r_{2}) = (4, 3.5)\), where the transmission probability nearly vanishes when $\vert\Gamma_{2}\vert > 1$. Analogous features can be seen for the third (\(S=3\)) and fourth (\(S=4\)) order super periodic delta potentials, as depicted in Fig. \ref{figure05}c and Fig. \ref{figure05}d, respectively. These figures again demonstrated that the transmission probability nearly vanished when the argument \(\Gamma_{3}\) and \(\Gamma_{4}\) exceeds the interval \([-1, 1]\).\\
\indent
Chebyshev polynomial of degree $N$ has $N$ roots in the interval $[-1,1]$. Hence from Eq. (\ref{eq5}), it is evident that $U_{N_{1}-1}(\Gamma_{1})$ has $N_{1}-1$ roots implying that Dirac comb of $N_{1}$ potentials has $N_{1}-1$ transmission resonances for $\Gamma_{1} \in [-1,1]$. From Eq. (\ref{eq17}), we can express $\Gamma_{2}$ as
\begin{equation}
    \Gamma_{2} = \vert M_{22}(k) \vert\cos[\tau-k\{(N_{1}-1)r_{1}-r_{2}\}]U_{N_{1}-1}(\Gamma_{1})-U_{N_{1}-2}(\Gamma_{1})\cos[k(N_{1}r_{1}-r_{2})].
\end{equation}
Let $\Gamma^{\star}_{1}=\Gamma_{1}$ ($k^{\star}_{(1)}=k$) is one of the roots of $U_{N_{1}-1}(\Gamma_{1})=0$, so from the above equation, the value of $\Gamma_{2}$ at $k=k^{\star}_{(1)}$ will be
\begin{equation}
    \Gamma_{2}(\Gamma^{\star}_{1})=-U_{N_{1}-2}(\Gamma^\star_{1})\cos[k^{\star}_{(1)}(N_{1}r_{1}-r_{2})]
\end{equation}
The subscript $(1)$ in $k^{\star}_{(1)}$ indicates that this transmission resonance is due to the SPP of order $S=1$. Generally, if a transmission resonance point is associated with the SPP of order $S$, it will be denoted as $k^{\star}_{(S)}$. As $U_{N_{2}-1}(\Gamma_{2})$ has $N_{2}-1$ roots for $\Gamma_{2}\in[-1,1]$. Hence, in the case of a periodic Dirac comb, with $N_{2}$ periodic counts, those energy points for which Dirac comb (periodic delta potential of periodic count $N_{1}$) had transmission resonances will split into a \textit{resonance band} containing $N_{2}-1$ resonance peaks. It should be noted that the $N_2-1$ peaks exclude the peak at $k = k^{\star}_{(1)}$, which is specifically associated with the transmission resonance of the Dirac comb itself. Consequently, the resonance band around $k^{\star}_{(1)}$ will encapsulate $N_2-1$ resonant peaks in addition to the peak at $k^{\star}_{(1)}$, a total of $(N_2-1)+1=N_{2}$ resonant peaks within this band.  Further, from Eq. (\ref{eq17}), $\Gamma_{3}$ can be expressed as
\begin{align}
    \Gamma_{3} &= \begin{aligned}[t]
        &\vert M_{22}(k) \vert \cos[\tau-k\{(N_{1}-1)r_{1}+(N_{2}-1)r_{2}-r_{3}\}]U_{N_{1}-1}(\Gamma_{1})U_{N_{2}-1}(\Gamma_{2}) \\
        &-U_{N_{1}-2}(\Gamma_{1})U_{N_{2}-1}(\Gamma_{2})\cos[k(N_{1}r_{1}+(N_{2}-1)r_{2}-r_{3})] \\
        &-U_{N_{2}-2}(\Gamma_{2})\cos[k(N_{2}r_{2}-r_{3})].
    \end{aligned}
    \label{eq20}
\end{align}
Again, at $\Gamma^\star_{1}=\Gamma_{1}$ $(k^{\star}_{(1)}=k)$, which is the root of $U_{N_{1}-1}(\Gamma_{1})$, the above equation becomes
\begin{align}
    \Gamma_{3}(\Gamma^{\star}_{1}) &= \begin{aligned}[t]
        &-U_{N_{1}-2}(\Gamma^\star_{1})U_{N_{2}-1}(\Gamma_{2}(\Gamma^{\star}_{1}))\cos[k^{\star}_{(1)}(N_{1}r_{1}+(N_{2}-1)r_{2}-r_{3})] \\
        &-U_{N_{2}-2}(\Gamma_{2}(\Gamma^{\star}_{1}))\cos[k^{\star}_{(1)}(N_{2}r_{2}-r_{3})].
    \end{aligned}
\end{align}
Now, let $\Gamma^\star_{2}=\Gamma_{2}$ ($k^{\star}_{(2)}=k$) is one of the roots of $U_{N_{2}-1}(\Gamma_{2})=0$, so from Eq. (\ref{eq20}), the value of $\Gamma_{3}$ at $k=k^{\star}_{(2)}$ will be
\begin{equation}
    \Gamma_{3}(\Gamma^{\star}_{2})=-U_{N_{2}-2}(\Gamma^\star_{2})\cos[k^{\star}_{(2)}(N_{2}r_{2}-r_{3})]
    \label{eq22}
\end{equation}
As $U_{N_{3}-1}(\Gamma_{3})$ has $N_{3}-1$ roots for $\Gamma_{3}=[-1,1]$. Therefore in the case of super periodic delta potential of order $3$ with periodic count $N_{3}$, those energy points for which Dirac comb had transmission resonances will split into a resonance band containing $N_{3}-1$ resonance peaks, excluding the peak at $k = k^{\star}_{(1)}$ which demonstrates transmission resonance due to the Dirac comb. Consequently, the resonance band hovered around $k^{\star}_{(1)}$ will feature a total of $(N_3 - 1) + 1 = N_3$ resonance peaks. Similarly, according to Eq. (\ref{eq22}), for energy points where the periodic Dirac comb exhibits transmission resonances, a resonance band will form, consisting of $N_3 - 1$ peaks, excluding the peak at $k = k^{\star}_{(2)}$, which displays transmission resonance due to the periodic Dirac comb (super periodic delta potential of order $2$ with periodic count $N_2$). In this scenario as well, the resonance band hovered around $k = k^{\star}_{(2)}$ will contain a total of $(N_3 - 1) + 1 = N_3$ resonance peaks.\\
\indent
In general, if $\Gamma^{\star}_{S-1}=\Gamma_{S-1}$ ($k^{\star}_{(S-1)}=k$) represents one of the roots of $U_{N_{S-1}-1}(\Gamma_{S-1})=0$ for $\Gamma_{S-1}\in[-1,1]$, then from Eq. (\ref{eq17}), $\Gamma_{S}$ at $k^{\star}_{(S-1)}=k$ is expressed through
\begin{equation}
\Gamma_{S}(\Gamma^{\star}_{S-1})=-U_{N_{S-1}-2}(\Gamma^\star_{S-1})\cos[k^{\star}_{(S-1)}(N_{S-1}s_{S-1}-s_{S})].
\label{eq23}
\end{equation}
Again. in general, for the super periodic delta potential of order $S$ with a super periodic count of $N_{S}$, the presence of $N_{S}-1$ roots for $\Gamma_{S}=[-1,1]$ in the function $U_{N_{S}-1}(\Gamma_{S})$ signifies that energy points where the super periodic delta potential of order $S-1$ with a super periodic count of $N_{S-1}$ exhibited transmission resonances will divide into a resonance band. This band will consist of $N_{S}-1$ peaks, excluding the resonance peak at $k=k^{\star}_{(S)}$. Therefore, around $k^{\star}_{(S-1)}$, there will be a total of $(N_{S}-1)+1=N_{S}$ resonance peaks. Moreover, as the super periodic count $N_S$ increases, the resolution of these resonance peaks becomes increasingly difficult, primarily due to the fixed range of $k$ values for which $|\Gamma_S| < 1$. This limited interval imposes constraints on the distinguishability of individual peaks, especially with the growing number of resonant peaks, which results in a higher peak density.\\
\indent
Fig. \ref{figure06} illustrates the formation of the resonance band for an SPP of order $S$, hovered around the resonance points of the preceding order $S-1$. Here, the potential parameter is set to $V=2$. Figs. \ref{figure06}a and \ref{figure06}b show the transmission profiles $T(2, 3, k)$ and $T(2, 3, 2, k)$ for super periodic delta potentials of orders $S=2$ and $S=3$ with super periodic distances $(r_{1}, r_{2})=(1.25, 2)$ and $(r_{1}, r_{2}, r_{3})=(1.25, 2, 6)$ respectively. In Fig. \ref{figure06}a, the transmission probability $T(2, 3, k)$ has two resonance points located at $k^{\star}_{(2)} = 2.779173, 3.106119$ and in Fig. \ref{figure06}b, the transmission probability $T(2, 3, 2, k)$ has two resonance points located at $k^{\star}_{(3)} = 2.937791, 3.106119$. These resonance points are marked with red grid lines on the $k$-axis in both figures. Considering the super periodic delta potential of order \(S = 2\), whose transmission profile is illustrated in Fig. \ref{figure06}a, as a \textit{unit cell} and periodically repeated with periodic counts \(N_{3} = 5, 10,\) and \(15\) with periodic distance \(r_{3} = 15\). The resulting transmission profiles \(T(2, 3, 5, k)\), \(T(2, 3, 10, k)\), and \(T(2, 3, 15, k)\) are depicted in Figs. \ref{figure06}c, \ref{figure06}e, and \ref{figure06}g, respectively. In Fig. \ref{figure06}c, the transmission profile \(T(2, 3, 5, k)\) exhibits a resonance band containing \(N_{3} = 5\) peaks in the proximity of the resonance points \(k^\star_{(2)} = 2.779173, 3.1069119\). Similarly, in Figs. \ref{figure06}e and \ref{figure06}g, the transmission profile $T(2, 3, 10, k)$ and $T(2, 3, 15, k)$ exhibits the resonance bands containing \(N_{3} = 10\) and \(15\) 
\begin{figure}[H]
\begin{center}
\includegraphics[scale=0.32]{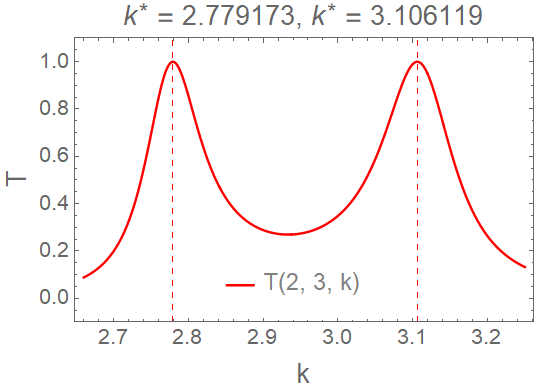} a
\includegraphics[scale=0.32]{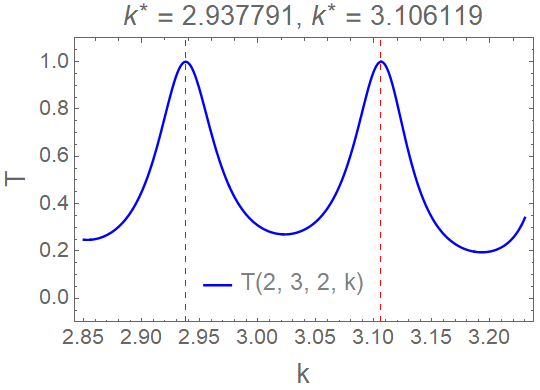} b \\
\includegraphics[scale=0.32]{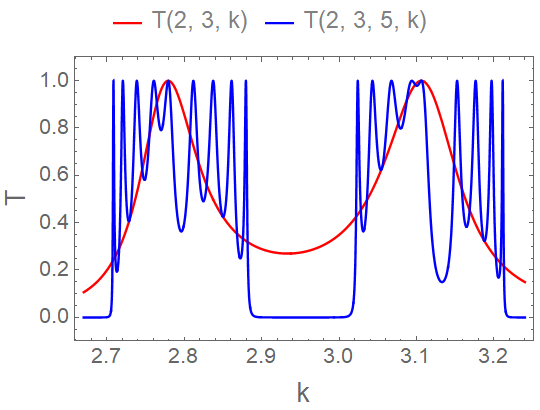} c 
\includegraphics[scale=0.32]{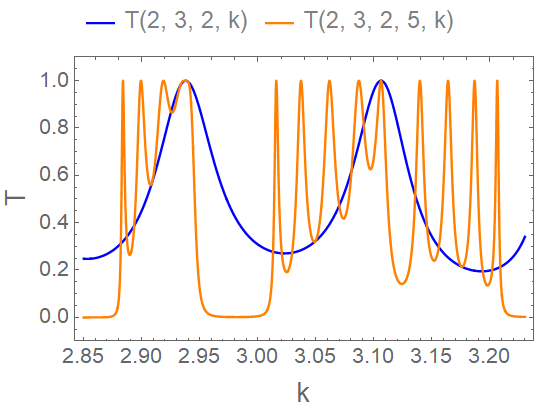} d \\
\includegraphics[scale=0.32]{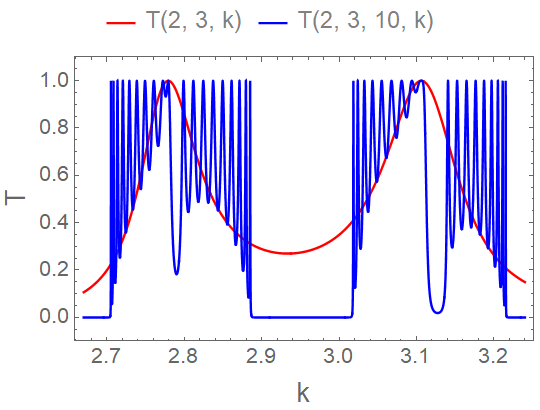}  e
\includegraphics[scale=0.32]{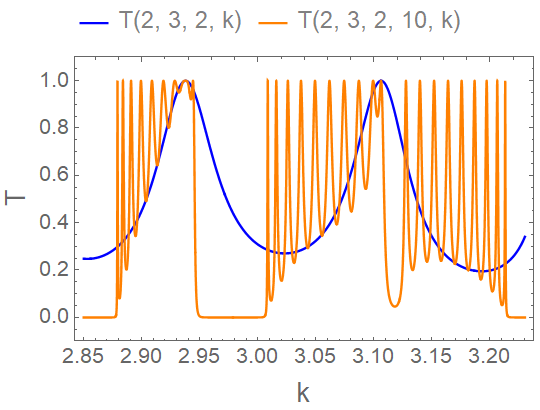} f \\
\includegraphics[scale=0.32]{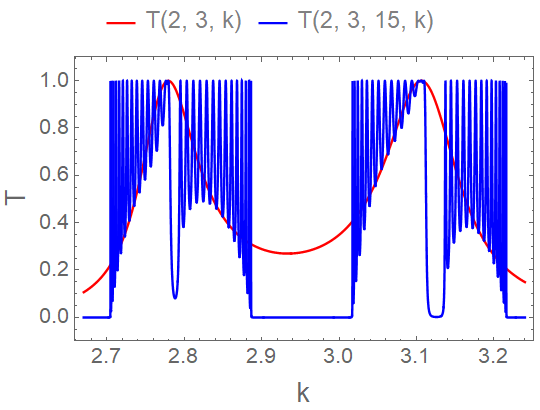}  g
\includegraphics[scale=0.32]{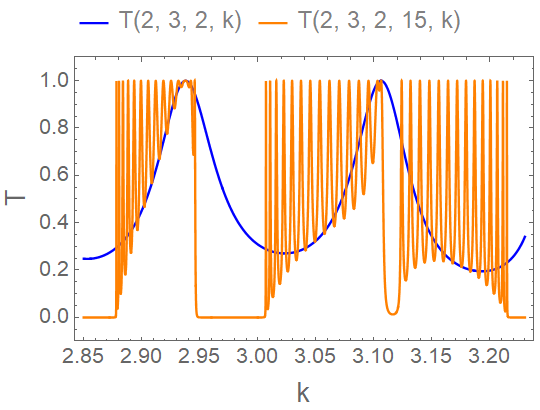} h
\caption{\it The figure illustrates the formation of resonance bands for SPP of order \(S\) around resonance points of order \(S-1\). Figs. (a) and (b) show transmission profiles \(T(2, 3, k)\) and \(T(2, 3, 2, k)\) for super periodic delta potentials of orders \(S=2\) and \(S=3\) with  super periodic distances \((r_{1}, r_{2}) = (1.25, 2)\) and \((r_{1}, r_{2}, r_{3}) = (1.25, 2, 6)\), respectively. Red grid lines mark the resonance points: \(k^\star_{(2)} = 2.779173, 3.106119\) in (a) and \(k^\star_{(3)} = 2.937791, 3.106119\) in (b). Figs. (c), (e), and (g) display transmission profiles \(T(2, 3, 5, k)\), \(T(2, 3, 10, k)\), and \(T(2, 3, 15, k)\) with periodic counts \(N_{3} = 5, 10, 15\) and super periodic distance \(r_{3} = 15\), showing \(N_{3}\) peaks forming a resonance band near \(k^\star_{(2)}\). Similarly, Figs. (d), (f), and (h) present transmission profiles \(T(2, 3, 2, 5, k)\), \(T(2, 3, 2, 10, k)\), and \(T(2, 3, 2, 15, k)\) with periodic counts \(N_{4} = 5, 10, 15\) and super periodic distance \(r_{4} = 15\), each showing \(N_{4}\) peaks forming the resonance band near \(k^\star_{(3)}\).}
\label{figure06}
\end{center}
\end{figure}
\noindent
peaks, respectively, can be seen around the same resonance points \(k^\star_{(2)}\). Next, Fig. \ref{figure06}b shows the transmission profile for the $T(2, 3, 2, k)$ having two resonance points. Again by considering this system as a \textit{unit cell} and periodically repeated with periodic counts $N_{4}=5, 10$ and $15$ at periodic distance $r_{4}=15$ then the resultant transmission profile $T(2, 3, 2, 5, k)$, $T(2, 3, 2, 10, k)$ and $T(2, 3, 2, 15, k)$ is shown in Figs. \ref{figure06}d, \ref{figure06}f and \ref{figure06}h respectively. In Fig. \ref{figure06}d, the transmission profile \(T(2, 3, 2, 5, k)\) shows a resonance band featuring \(N_{3} = 5\) peaks close to the resonance points \(k^\star_{(3)} = 2.937791, 3.1069119\). Similarly, Figs. \ref{figure06}f and \ref{figure06}h depict the transmission profiles \(T(2, 3, 2, 10, k)\) and \(T(2, 3, 2, 15, k)\), which display resonance bands with \(N_{3} = 10\) and \(N_{3} = 15\) peaks, respectively, near the same resonance points \(k^\star_{(3)}\). In general, the resonance band containing $N_{S}$ resonance peaks in the transmission profile of a potential system of order \( S \) can be observed in the vicinity of the resonance point found in the transmission profile of the potential system of preceding order \( S-1 \).
\subsection{Transmission through CDC-\texorpdfstring{$\rho_{N}$}{N} system}
\begin{figure}[H]
\begin{center}
\includegraphics[scale=0.468]{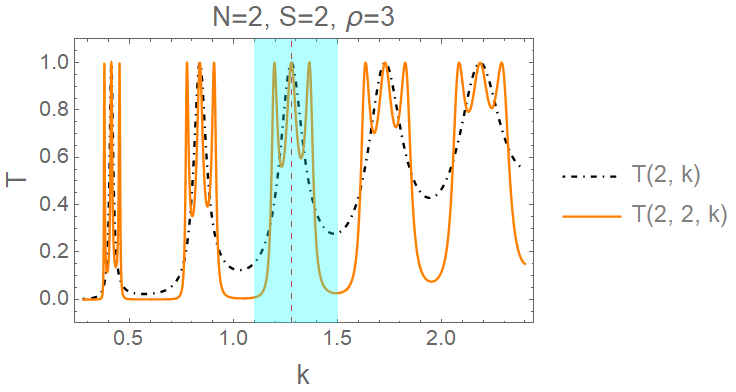} a \\
\includegraphics[scale=0.468]{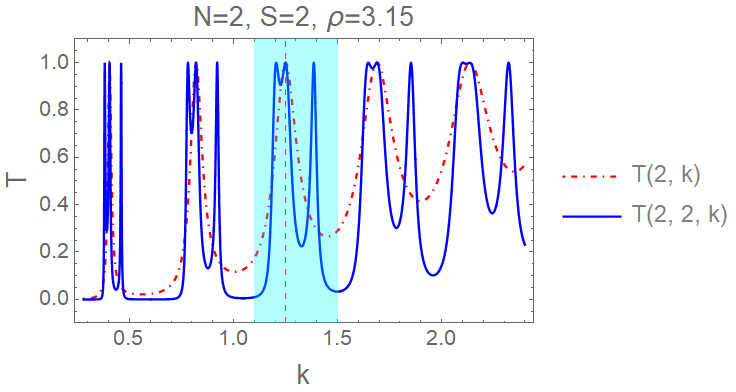} b \\
\includegraphics[scale=0.468]{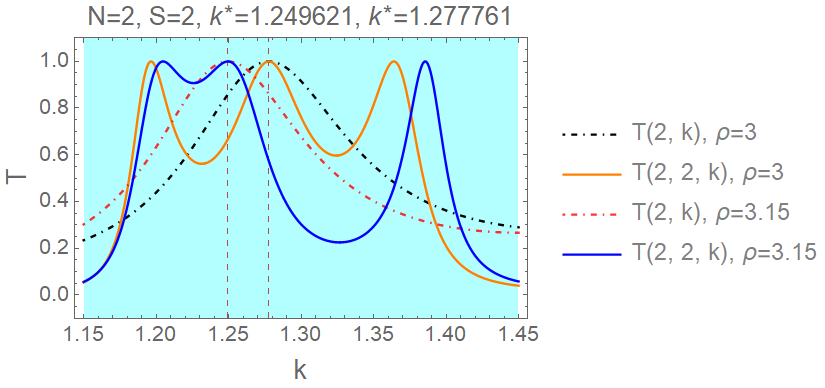} c 
\caption{\it 
Transmission probability \(T(2, k)\) and \(T(2, 2, k)\) through the CDC-$\rho_{2}$ system at stage \(S=2\) for scaling parameters (a) \(\rho=3\) and (b) \(\rho=3.15\). The potential parameters are maintained at \(V=1\) and \(L=20\). This figure demonstrates the alterations in the transmission profile as the system transitions from periodic to super periodic configurations. A magnified view of the transmission profiles within the \(k\) range of 1.1 to 1.5, highlighted in cyan, is shown in Fig. (c).
}
\label{figure07}
\end{center}
\end{figure}

\begin{figure}[H]
\begin{center}
\includegraphics[scale=0.375]{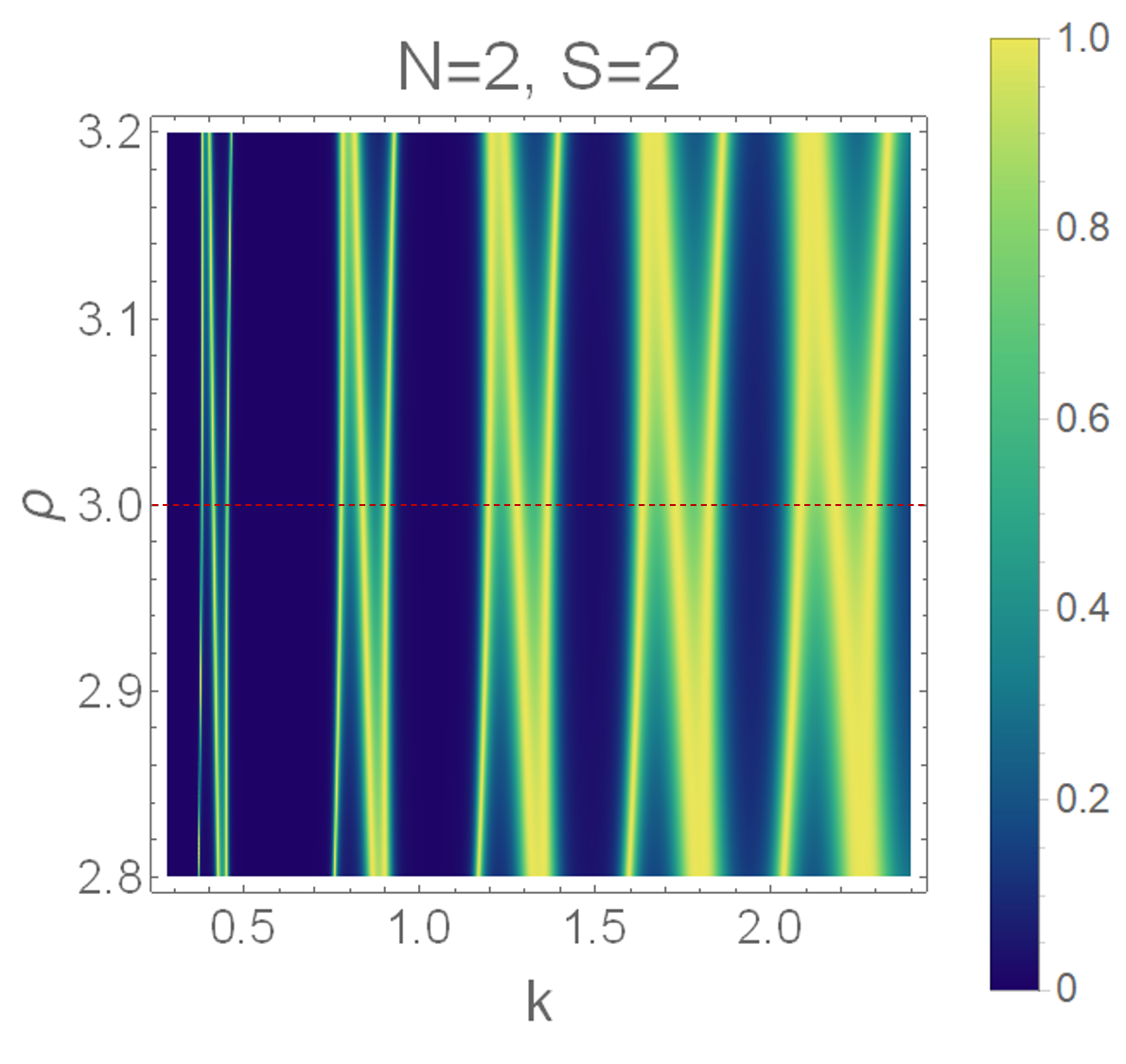} a 
\includegraphics[scale=0.375]{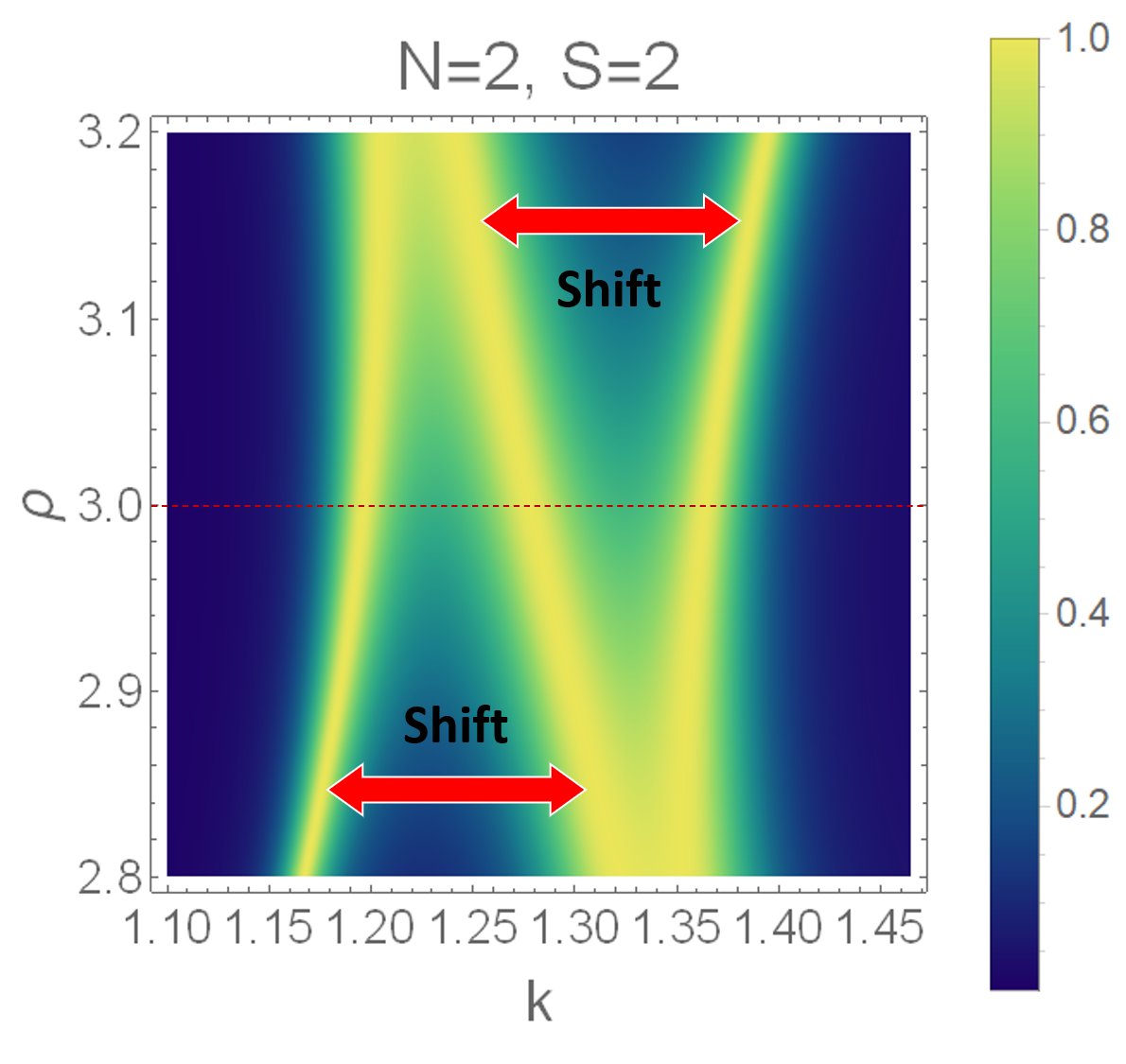} b \\
\includegraphics[scale=0.375]{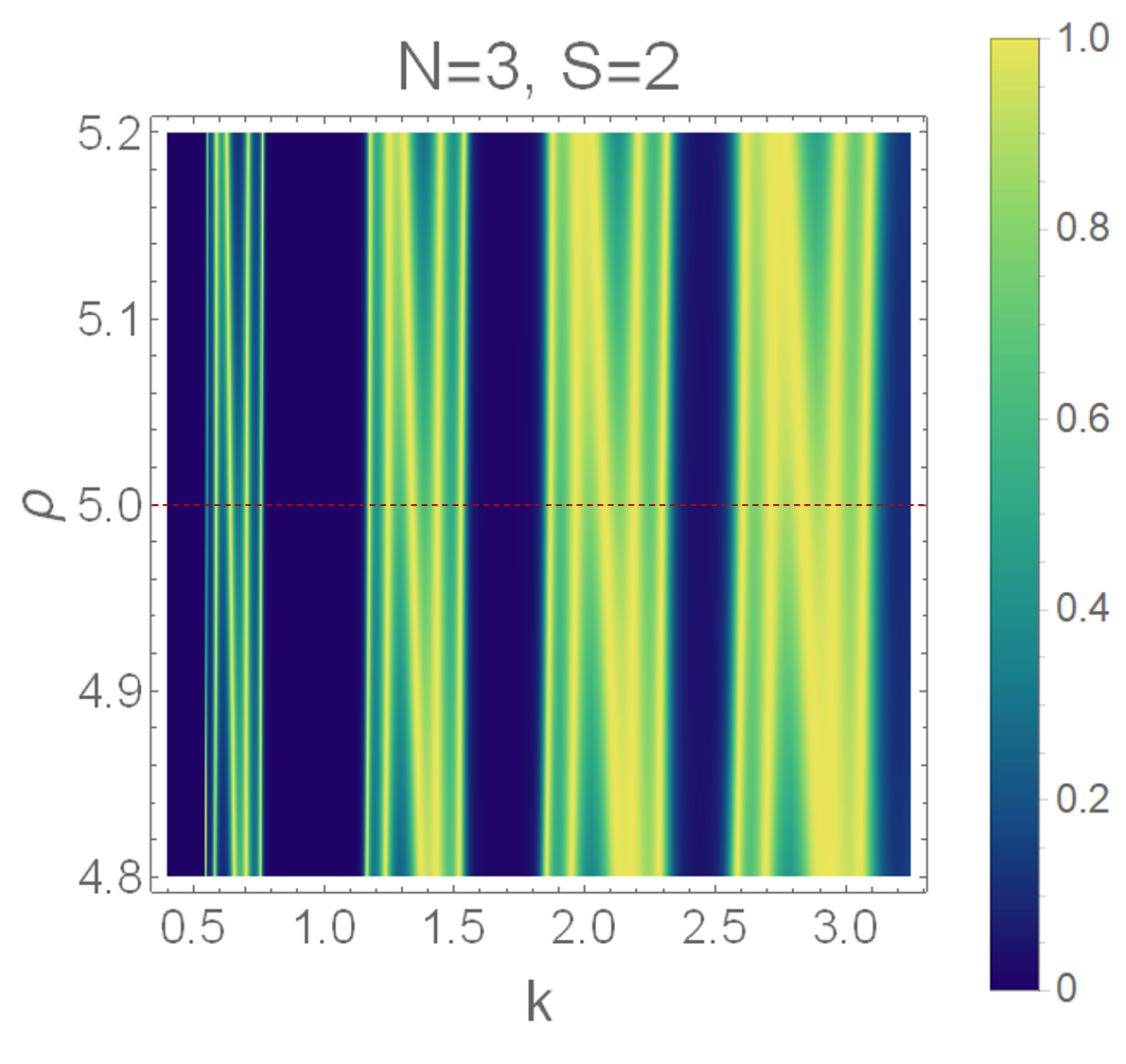} c 
\includegraphics[scale=0.375]{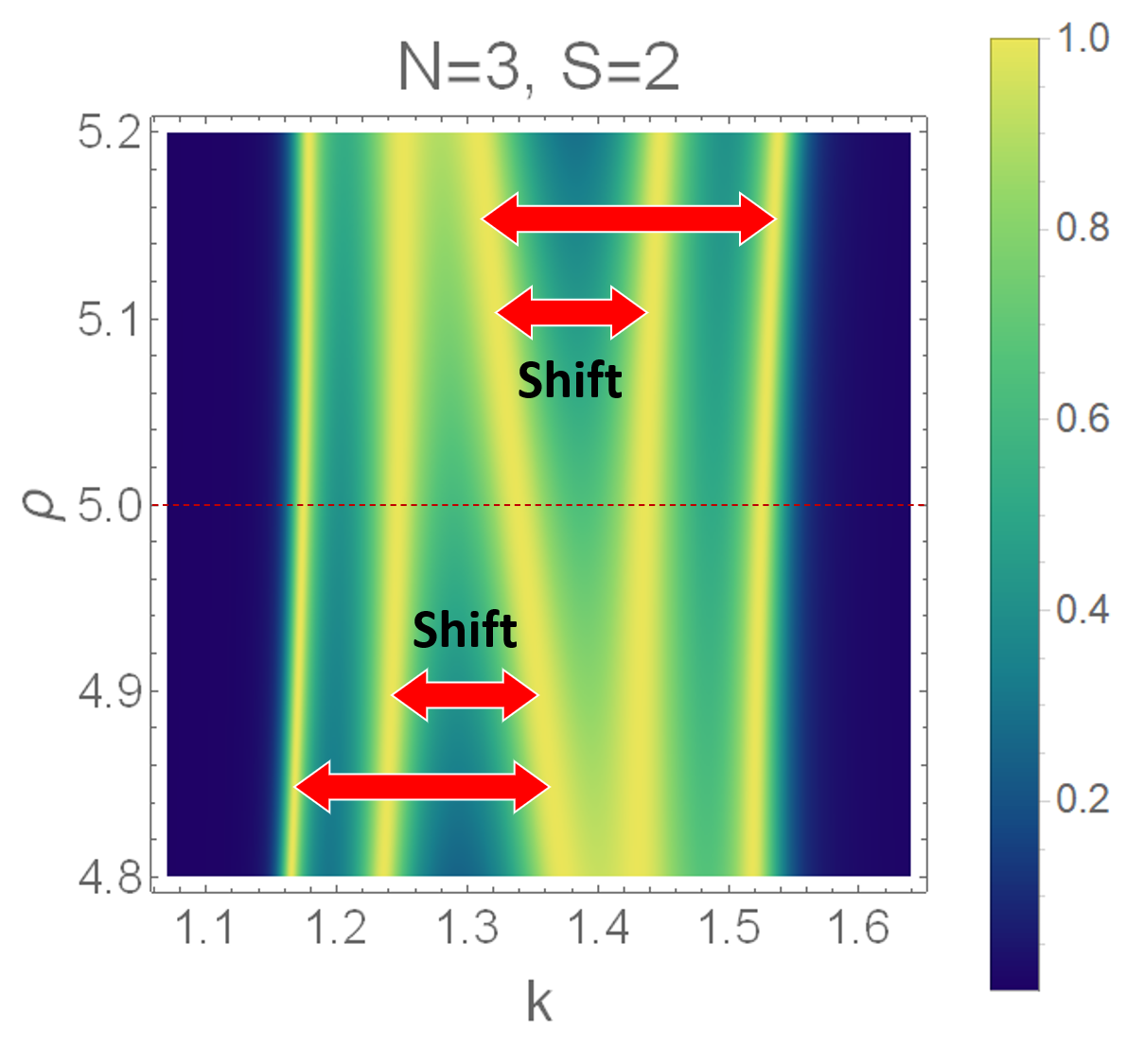} d
\caption{\it Density plots showing the transmission probability in the \(\rho-k\) plane for stage \(S=2\) of the (a) CDC-$\rho_{2}$ and (c) CDC-$\rho_{3}$ systems. The potential parameters are the same as of Fig. $\ref{figure07}$. The color bar legend indicates the variation of \(T(k)\), where \(T(k)=0\) corresponds to blue and \(T(k)=1\) corresponds to yellow. Figs. (b) and (d) present magnified views of Figs. (a) and (c) for the \(k\) ranges \(1.10\) to \(1.452\) and \(1.08\) to \(1.64\), respectively. Red arrows in these magnified figures indicate shifts in the transmission resonances from the central resonance peak. The red and blue shifts of the transmission resonances are observed for \(\rho > \rho_{p}\) and \(\rho < \rho_{p}\).
}
\label{figure08}
\end{center}
\end{figure}

\begin{figure}[H]
\begin{center}
\includegraphics[scale=0.395]{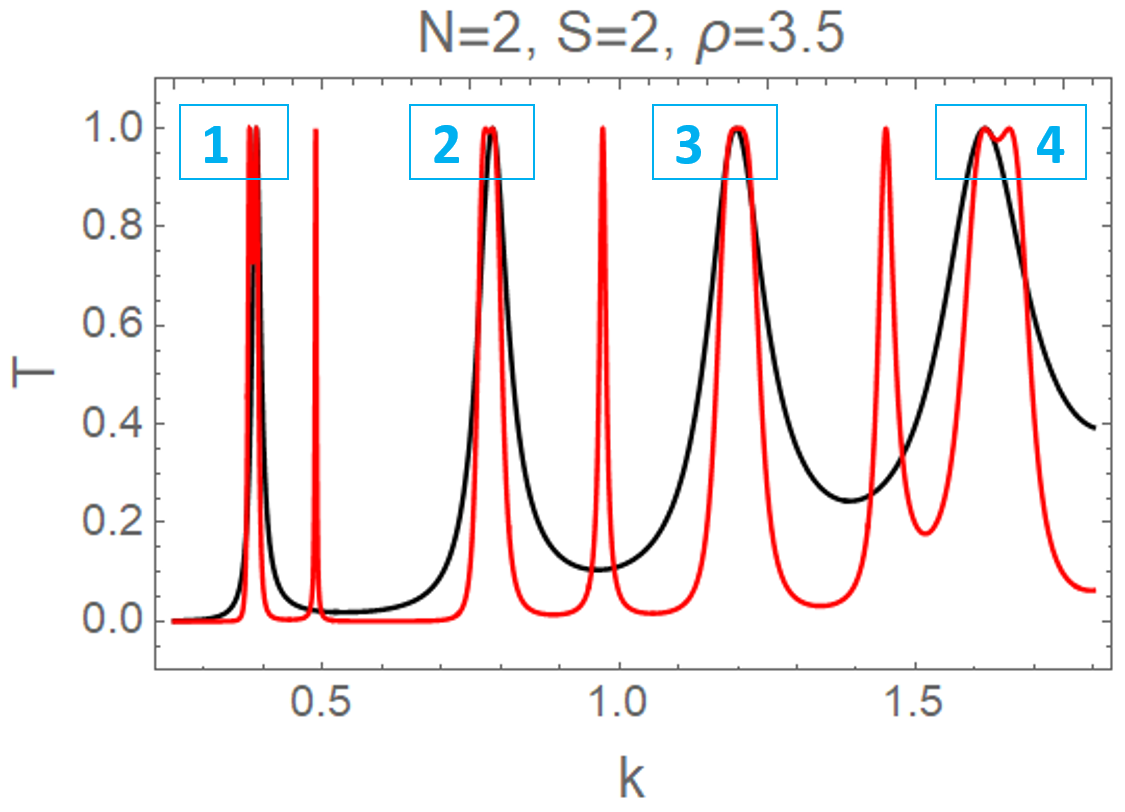} a
\includegraphics[scale=0.395]{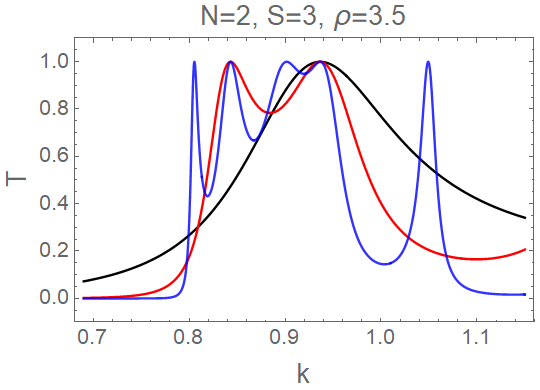} b \\
\includegraphics[scale=0.395]{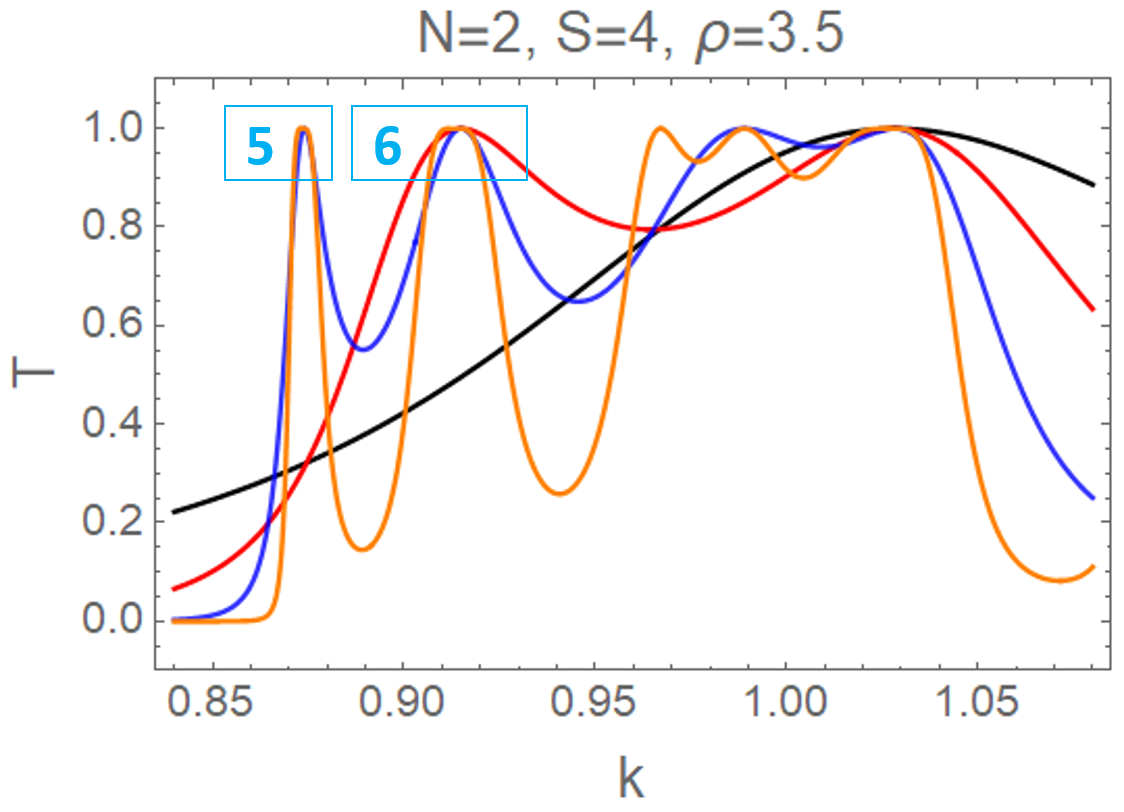} c
\includegraphics[scale=0.414]{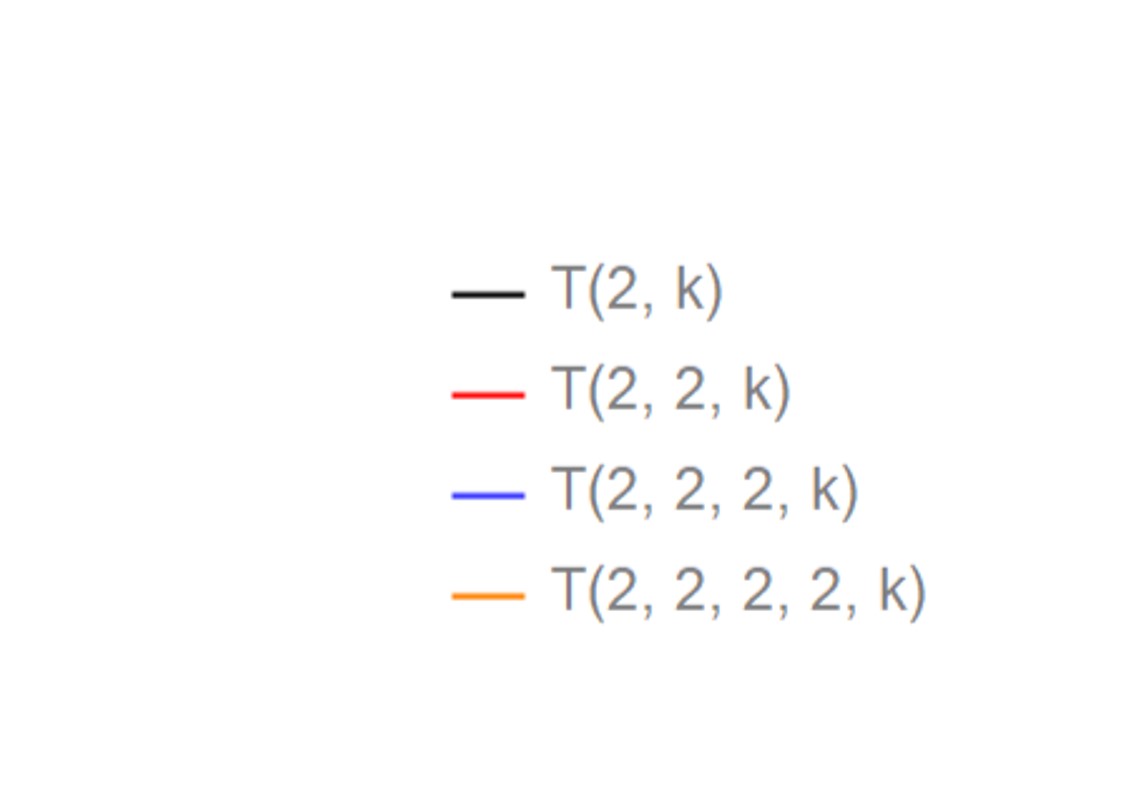} 
\caption{\it Figure illustrating the transmission probability for the CDC-$\rho_{2}$ system at various stages: (a) \(T(2, k)\) and \(T(2, 2, k)\) for stage \(S=2\), (b) \(T(2, k)\), \(T(2, 2, k)\) and \(T(2, 2, 2, k)\) for stage \(S=3\) and (c) \(T(2, k)\), \(T(2, 2, k)\), \(T(2, 2, 2, k)\) and \(T(2, 2, 2, 2, k)\) for stage \(S=4\). Here, the potential parameters are $V=1$, $L=20$ and $\rho=3.5$. A magnified view of the transmission resonance peaks (pair of peaks), labeled as $1$, $2$, $3$, $4$, $5$ and $6$, is presented in Fig. $\ref{figure10}$.} 
\label{figure09}
\end{center}
\end{figure}

\begin{figure}[h! tbp]
\begin{center}
\includegraphics[scale=0.395]{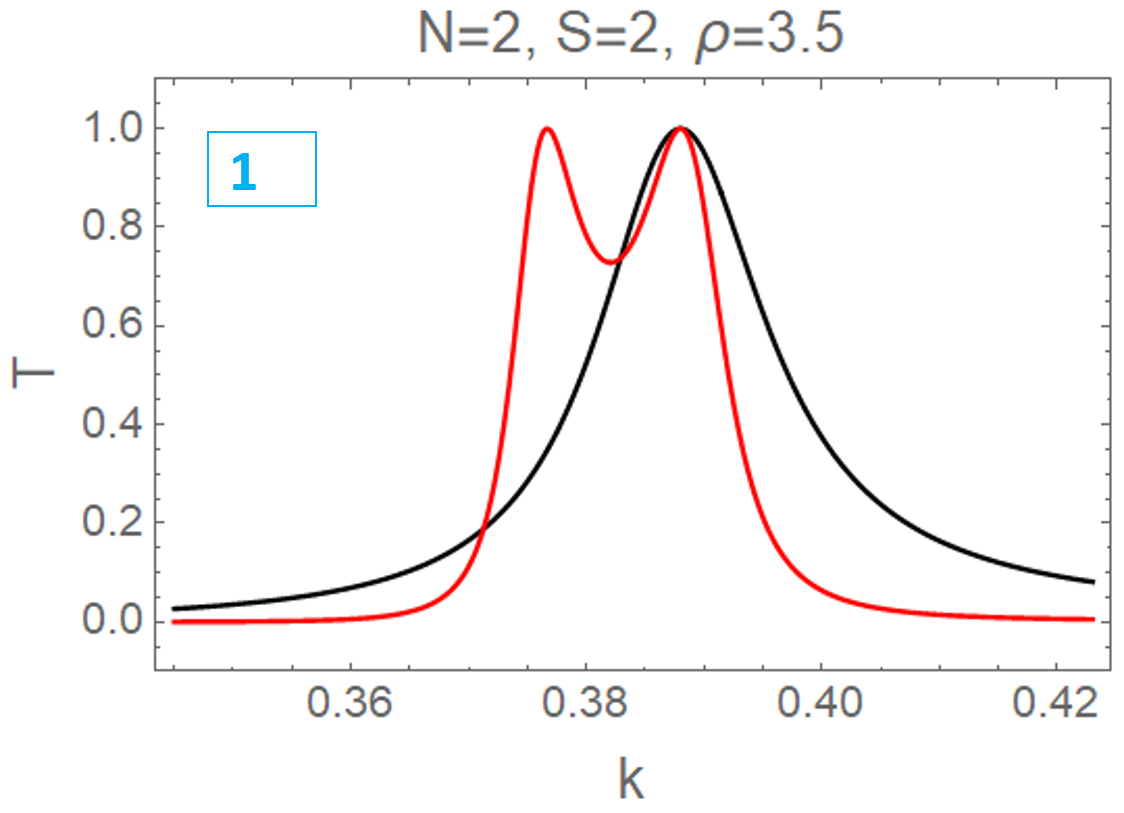} a
\includegraphics[scale=0.395]{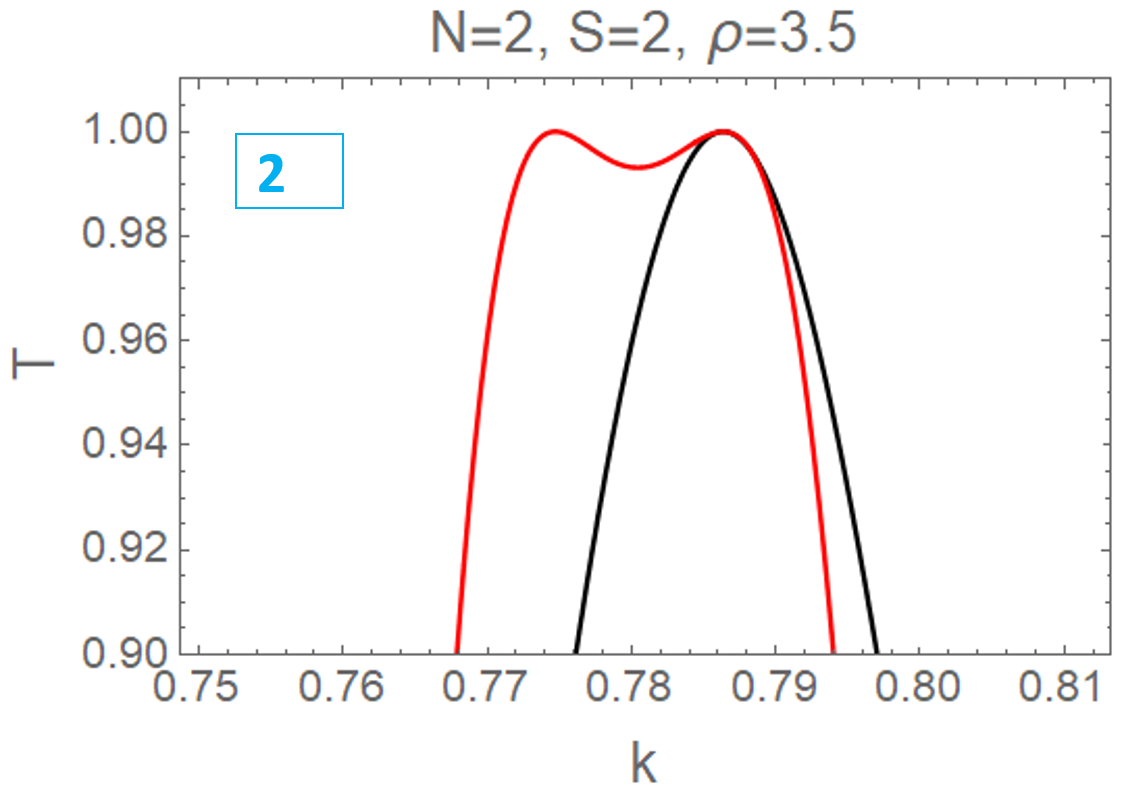} b \\
\includegraphics[scale=0.395]{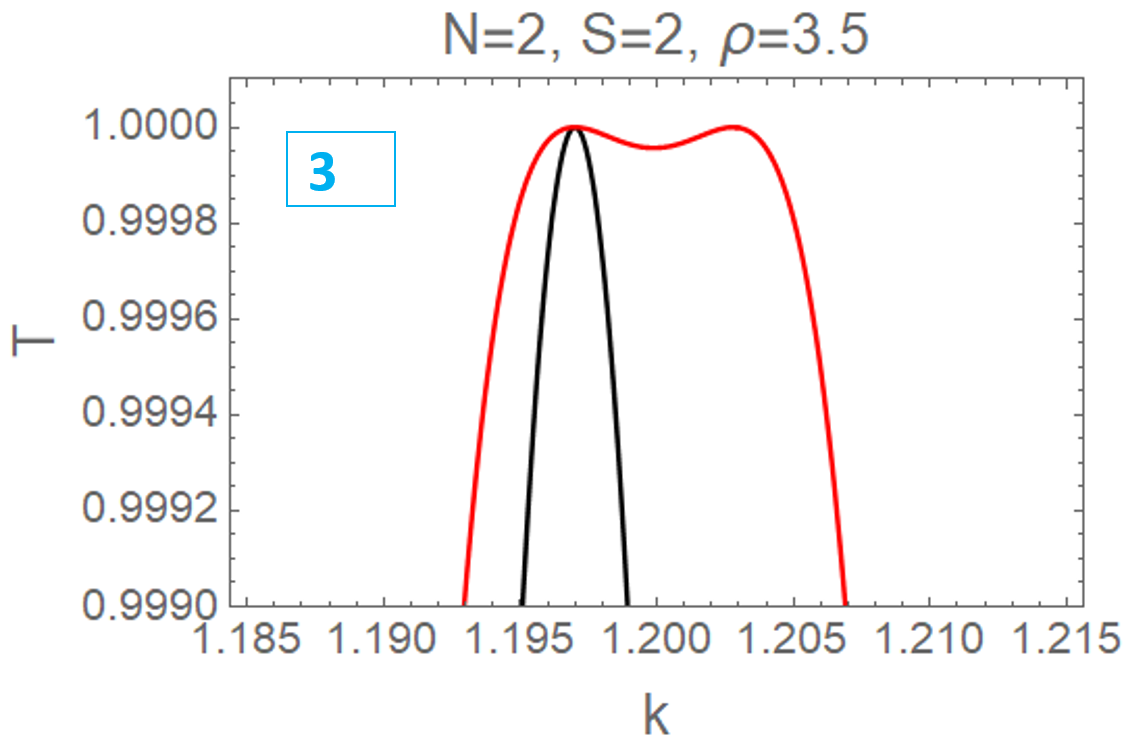} c
\includegraphics[scale=0.395]{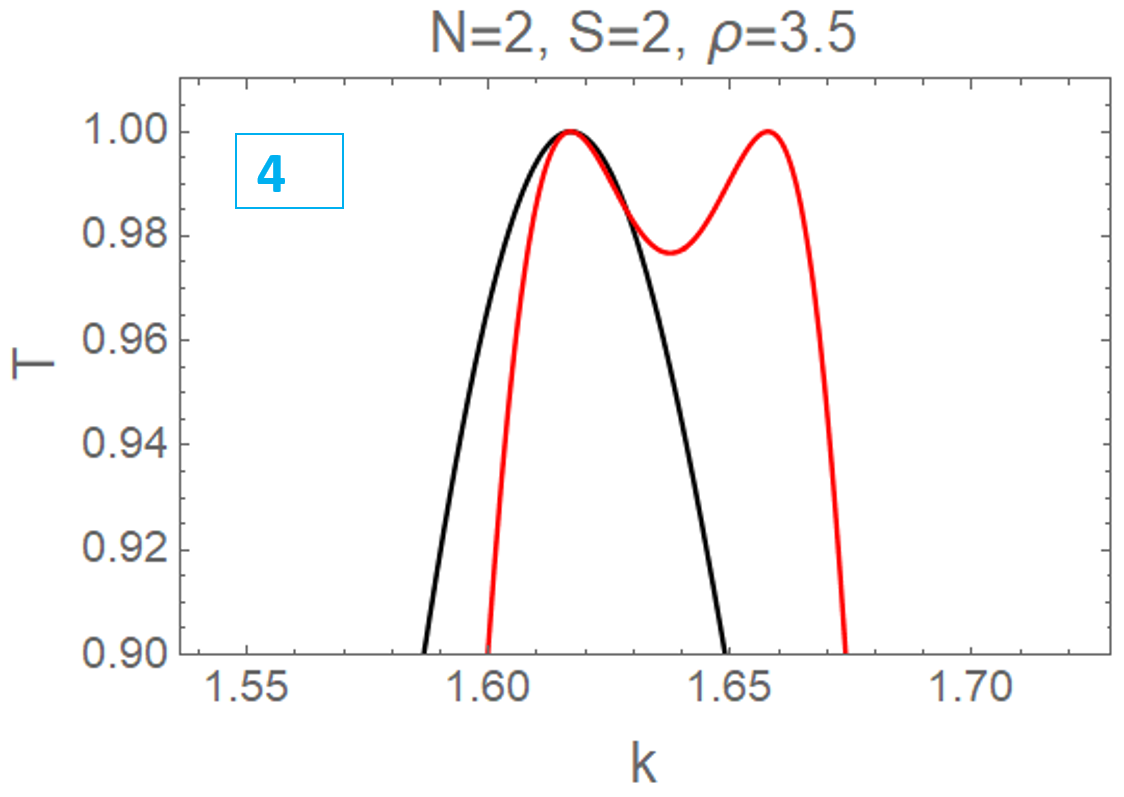} d \\
\includegraphics[scale=0.395]{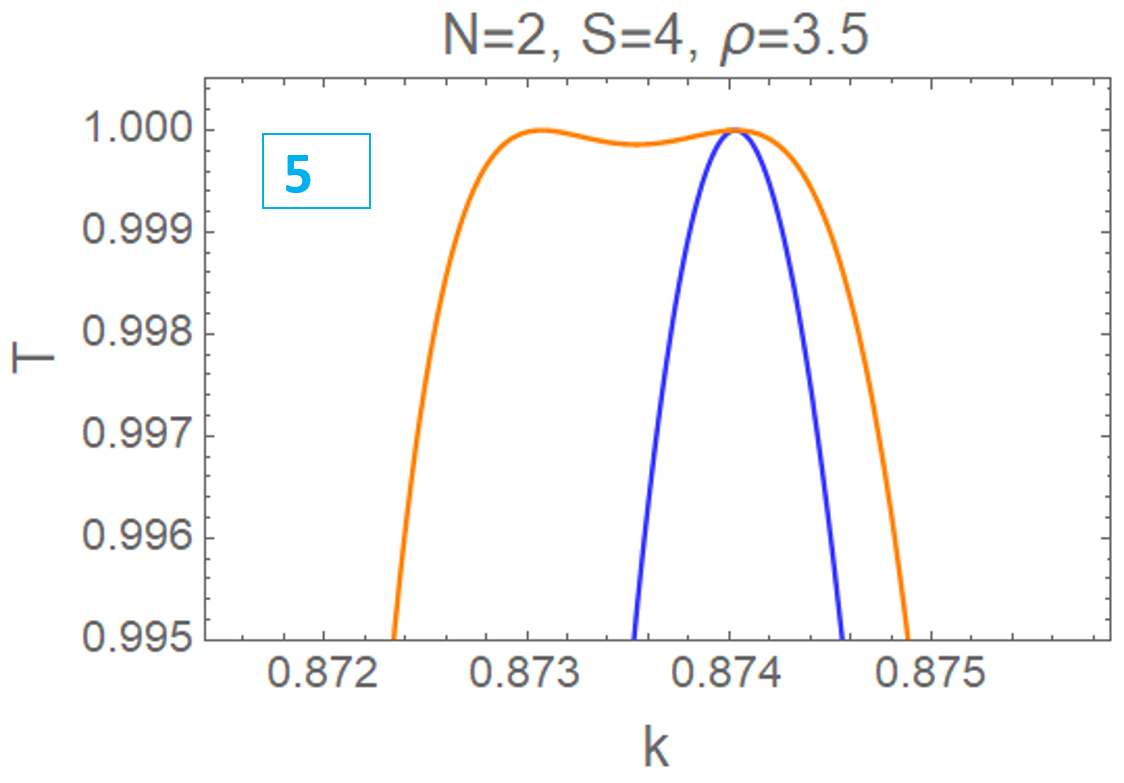} a
\includegraphics[scale=0.395]{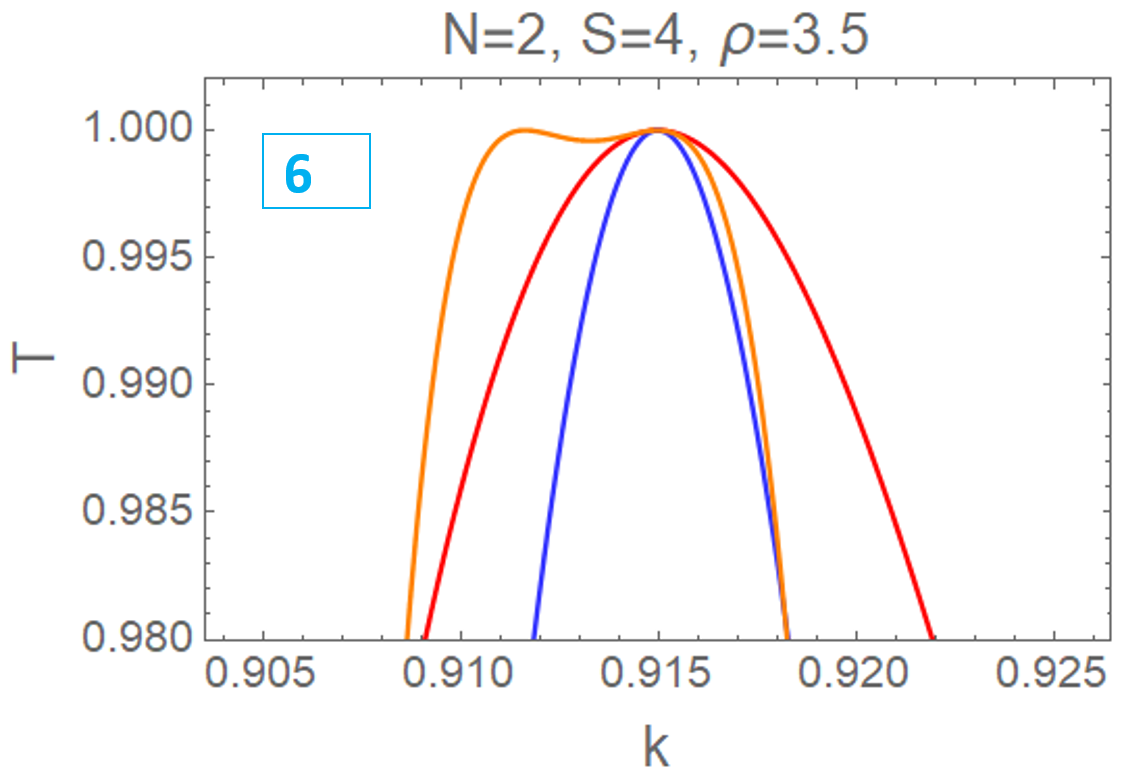} b 
\caption{\it A magnified view of the transmission resonance peaks of Fig. $\ref{figure09}$, labeled $1$, $2$, $3$, $4$, $5$ and $6$, is shown. Two resonance peaks are observed around each transmission resonance point in all the figures.}
\label{figure10}
\end{center}
\end{figure}

\begin{figure}[H]
\begin{center}
\includegraphics[scale=0.395]{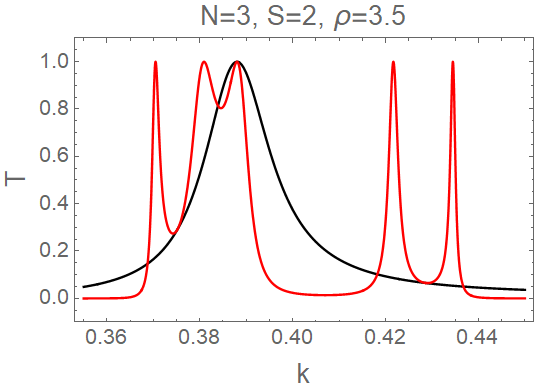} a
\includegraphics[scale=0.395]{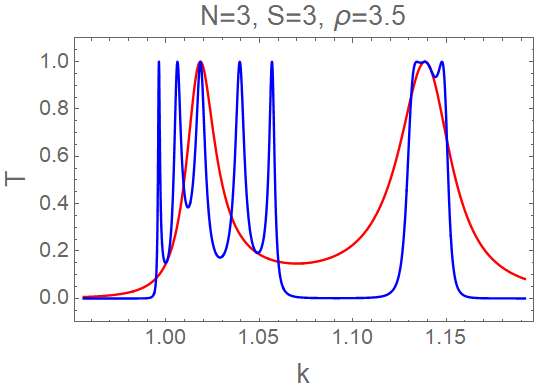} b \\
\includegraphics[scale=0.395]{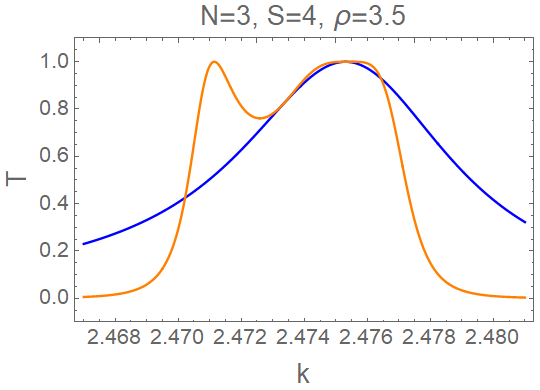} c
\includegraphics[scale=0.414]{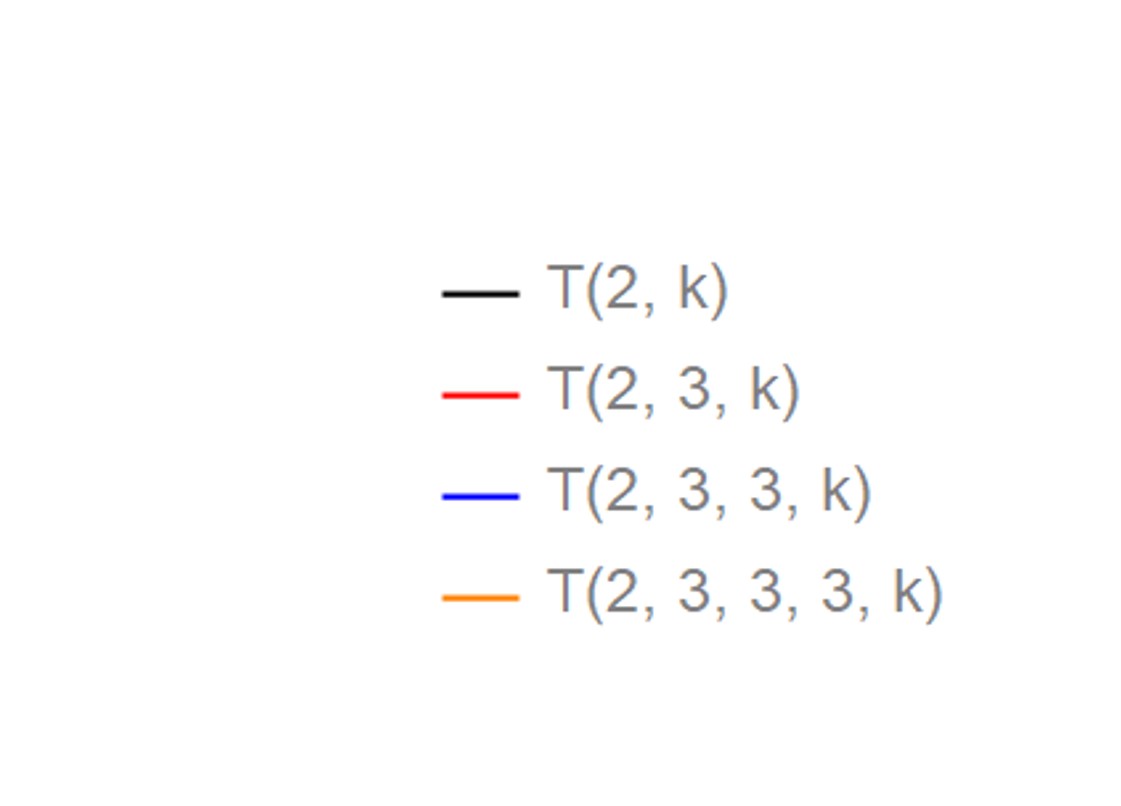} 
\caption{\it Illustration of the transmission probability for the CDC-$\rho_{3}$ system at various stages: (a) \(T(2, k)\) and \(T(2, 3, k)\) for stage \(S=2\); (b) \(T(2, k)\), \(T(2, 3, k)\), and \(T(2, 3, 3, k)\) for stage \(S=3\) and (c) \(T(2, k)\), \(T(2, 2, k)\), and \(T(2, 3, 3, 3, k)\) for stage \(S=4\). The potential parameters are set to \(V=1\), \(L=50\) and $\rho=3.5$. According to the definition of CDC$\rho_{3}$ system, \(N_{1}=2\) and \(N_{2, 3, \ldots, S} = 3\), implying that the transmission probability will exhibit three peaks at each resonance point of the transmission through the preceding order potential system. Here, Fig. (a), \(T(2, 3, k)\) (red curve) reveals three peaks at the resonance point of \(T(2, k)\) (black curve). Similarly, in (b), \(T(2, 3, 3, k)\) (blue curve) displays three resonance peaks around the resonance point of \(T(2, 3, k)\) (red curve). This pattern also holds for the next stage $S=4$ in Fig. (c) as well.}
\label{figure11}
\end{center}
\end{figure}

We have previously discussed the transmission characteristics of the SPP system, focusing on transmission resonances and the formation of band structures. In this subsection, we will analyze the transmission probability through the super periodic Cantor structured Dirac comb potential system. Firstly, we will analyze the changes in the transmission profile as the potential system transitions from a periodic to a super periodic geometry within the framework of the CDC-$\rho_{N}$ system.\\
\indent
The plots in Figs. \ref{figure07}a and \ref{figure07}b depict the transmission probability $T(2, k)$ and $T(2, 2, k)$ through the CDC-$\rho_{2}$ system at stage $S=2$ for scaling parameters $\rho=3$ and $\rho=3.15$ respectively. In these plots, the potential parameters are fixed at $V=1$ and $L=20$. From the geometry depicted in Fig. \ref{figure_01}, it is evident that for $\rho=3$, the stage $S=2$ of the CDC-$\rho_{2}$ system represents a periodic delta potential rather than a super periodic delta potential. This is because, in this case, the super periodic parameters $(N_{1}, N_{2}, r_{1}, r_{2}) = (2, 2, L/3, 2L/3)$ can be redefined by the periodic parameters $(N_{1}, r_{1}) = (4, L/3)$. Thus, for $\rho=3$, the CDC-$\rho_{2}$ system of stage $S=2$ is identified as a periodic delta potential with a periodic count of $N_{1}=4$ and a periodic distance of $r_{1}=L/3$. Notably, the periodicity of the delta potential at stage $S=2$ is disrupted when $\rho < 3$ or $\rho > 3$. The behaviour of $T(2, k)$ and $T(2, 2, k)$ for $\rho>3$ ($\rho=3.15$) is shown in Fig. \ref{figure07}b.\\
\indent
The Chebyshev polynomial of the second kind, $U_{N}(\Gamma)$, of degree $N$, has $N$ roots within the interval $\Gamma \in [-1, 1]$. Consequently, $U_{N_{1}-1}(\Gamma_{1})$ has $N_{1}-1$ roots for $\Gamma_{1} \in [-1, 1]$. Therefore, $T(N_{1}=2, k)$ features $2-1=1$ resonance peak for $\Gamma_{1} \in [-1, 1]$, as indicated by the black dot-dashed curve in Fig. \ref{figure07}a. For $\rho=3$, $T(N_{1}=2, N_{2}=2, k)$ identified as $T(N_{1}=4, k)$, resulting in $4-1=3$ resonance peaks in the set of resonance band for the new $\Gamma_{1} \in [-1, 1]$, depicted by orange curve in Fig. \ref{figure07}a. The width of this resonance band increases with $k$. It is evident from the figure that the resonance points of $T(2, k)$ coincide with the central peak resonances of $T(2, 2, k)$, while the remaining two peaks in each resonance band are symmetrically positioned on either side of this central peak. Fig. \ref{figure07}b illustrates the transmission probabilities \( T(2, k) \) and \( T(2, 2, k) \) for \(\rho = 3.15\) using the same potential parameters as in Fig. \ref{figure07}a. A zoomed view of both plots within the \(k\) range of $1.1$ to $1.5$, highlighted in cyan color, is shown in Fig. \ref{figure07}c. It is evident here that the transmission resonance point of \( T(2, k) \) for \(\rho = 3\) at \( k^\star = 1.277761 \) is redshifted to the resonance point at \( k^\star = 1.249261\) when \(\rho = 3.15\). Accordingly, the central peak resonance
\noindent
point of \( T(2, 2, k) \) for \(\rho = 3.15\) exhibits a redshift. Furthermore, the symmetry in the positions of the side peaks relative to the central peak in the resonance band, observed for \(\rho=3\) (periodic structure), is altered for \(\rho=3.15\) (super periodic structure). In this case, the left and right side peaks exhibit a blueshift. Consequently, \(\Delta k_{L}(\rho=3) > \Delta k_{L}(\rho=3.15)\) and \(\Delta k_{R}(\rho=3) < \Delta k_{R}(\rho=3.15)\), where \(\Delta k_{L}\) and \(\Delta k_{R}\) represent the ranges of $k$ from the central and left peak, and the central and right peak, respectively. As a result, the central resonance point with the nearest peak (left side peak) appears as the splitting of the peaks in each band according to the count of super periodicity, showcasing an important feature of super periodicity. Additionally, changes in the height of the valleys between the resonance points are observed when the system transitions from a periodic to a super periodic structure.\\
\indent
To present the more intricate and rich features of the tunneling profile for the case when stage CDC-\(\rho_{2}\) and CDC-\(\rho_{3}\) system of stage $S=2$ undergo a transition from periodic to super periodic structure, Fig. \ref{figure08} offers density plots in the \(\rho-k\) plane, illustrating this transition. The potential parameters are the same as those in Fig. \ref{figure07}. As mentioned earlier, the scaling parameter \(\rho=3\) results in a periodic delta potential system for \(N=2\) at stage \(S=2\). Likewise, \(\rho=5\) forms a periodic delta potential system for \(N=3\) at the stage. These density plots depict the transmission profile behavior in the \(\rho-k\) plane for two conditions: when \(\rho > \rho_{p}\) and when \(\rho < \rho_{p}\). Here, \(\rho_{p}=2N-1\) represents the critical value of \(\rho\) that forms the periodic structure of the delta potential for a given $N$ at the second stage ($S=2$). \\
\indent 
As discussed earlier, the second stage of the CDC-$\rho_{2}$ system with $\rho_{p}=3$ features a periodic delta potential, resulting in a transmission profile with bands containing three resonance peaks, as shown in Fig. \ref{figure07}a. Similarly, the second stage of the CDC-$\rho_{3}$ system with $\rho_{p}=5$ exhibits a periodic delta potential, resulting a transmission profile with bands that contain five resonance peaks, as illustrated in Fig. \ref{figure08}c. This occurs because for $\rho_{p}=5$, $T(N_{1}=2, N_{2}=3, k)$ corresponds to $T(N_{1}=6, k)$, leading to $6-1=5$ resonance peaks within the resonance band. In these density plots, the red dashed line indicates the critical value of the scaling parameter $\rho_{p}$ for the given $N$. In Fig. \ref{figure08}a, the locus of transmission resonance peaks, i.e. $T(2, 2, k)=1$, exhibits opposite behaviors for $\rho > \rho_{p} = 3$ and $\rho < \rho_{p} = 3$. Specifically, the central resonance peak at $\rho_{p} = 3$ undergoes a red shift for $\rho > \rho_{p}$ and a blue shift for $\rho < \rho_{p}$. Consequently, $\Delta k_{L}(\rho > \rho_{p}) < \Delta k_{L}(\rho < \rho_{p})$ and $\Delta k_{R}(\rho > \rho_{p}) > \Delta k_{R}(\rho < \rho_{p})$. To provide a clear view of the shift of the locus of $T(2, 2, k)=1$ in the $\rho-k$ plane, Fig. \ref{figure08}b presents a magnified view of Fig. \ref{figure08}a, focusing on the $k$ range from $1.10$ to $1.452$. A similar type of transmission feature is observed in Fig. \ref{figure08}c, where the central resonance peak at $\rho_{p} = 5$ undergoes a red shift for $\rho > \rho_{p}$ and a blue shift for $\rho < \rho_{p}$ Consequently, the same variation in $\Delta k_{L}$ and $\Delta k_{R}$ is observed above and below the critical value $\rho_{p}$. Again, for enhanced clarity in the shift of the locus of $T(2, 2, k)=1$, Fig. \ref{figure08}d presents a magnified view of Fig. \ref{figure08}c, focusing on the $k$ range from $1.08$ to $1.64$. These density plots show the changes in the transmission profile as the periodic delta potential transforms into a super periodic delta potential. It is evident that the central peak shifts closer to its neighboring peaks, moving left (redshift) for $\rho > \rho_{p}$ and moving right (blue shift) for $\rho < \rho_{p}$. This behavior results in the formation of a band-like feature containing $N_{2}$ peaks, due to the system being super periodically repeated $N_{2}$ times.\\
\indent
For a given stage $S$, the Cantor-structured Dirac comb potential allowed the super periodicity of the delta potential up to an order $S$. Therefore, the transmission probability
\noindent
for each order can be evaluated. Fig. \ref{figure09} illustrates the transmission profile for the CDC-$\rho_{2}$ system of stage $S=2$, $3$ and $4$ providing a comprehensive view of its resonance behavior. Here the potential parameter is taken as $V=1$, $L=20$ and scaling parameter $\rho=3.5$. For stage $S = 2$, super periodicity of up to order 2 is achievable, with the corresponding transmission probabilities $T(2, k)$ and $T(2, 2, k)$ depicted in Fig. \ref{figure09}a. Likewise, for stage $S = 3$, super periodicity extends to order $3$, and the transmission probabilities $T(2, k)$, $T(2, 2, k)$, and $T(2, 2, 2, k)$ are illustrated in Fig. \ref{figure09}b. In the similar fashion, for stage $S = 4$, super periodicity reaches up to order $4$, with the transmission probabilities $T(2, k)$,
\noindent
$T(2, 2, k)$, $T(2, 2, 2, k)$, and $T(2, 2, 2, 2, k)$ illustrated in Fig. \ref{figure09}c. Each stage progressively reveals higher orders of periodicity and their corresponding transmission behavior. The definition of the CDC-$\rho_{2}$ system says that for each order the super periodic count is $N_{1, 2, 3, \ldots, S} = 2$. As a result, the transmission probability shows two peaks at every resonance point of the transmission through the preceding order potential system, as demonstrated in Fig. \ref{figure09}. To provide a clearer view of the two peaks at the resonance points, a magnified display of the peaks labeled $1$, $2$, $3$, $4$, $5$, and $6$ is shown in Fig. \ref{figure10}. Next, Fig. \ref{figure11} illustrates the transmission profile for the CDC-$\rho_{3}$ system at various stages, $S=2$, $S=3$, and $S=4$. The transmission probabilities $T(2, k)$ and $T(2, 3, k)$ for stage $S=2$, $T(2, 3, k)$ and $T(2, 3, 3, k)$ for stage $S=3$, and $T(2, 3, 3, k)$ and $T(2, 3, 3, 3, k)$ for stage $S=4$ are displayed in Figs. \ref{figure11}a, \ref{figure11}b, and \ref{figure11}c, respectively. These figures illustrate the splitting of the transmission peaks into three distinct peaks at the resonance points of the transmission through the preceding order potential system. The appearance of three peaks at each resonance point arises from the fact that, in the CDC-$\rho_3$ system, $N_1 = 2$ and $N_{2, 3, 4, \ldots, S} = 3$, implying that all the super periodic count except first periodic count is three. 
\noindent

\begin{figure}[H]
\begin{center}
\includegraphics[scale=0.327]{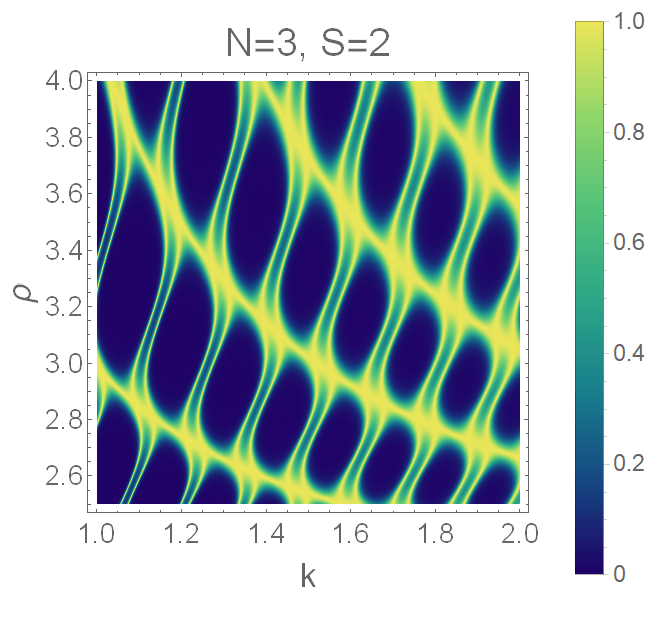} a
\includegraphics[scale=0.327]{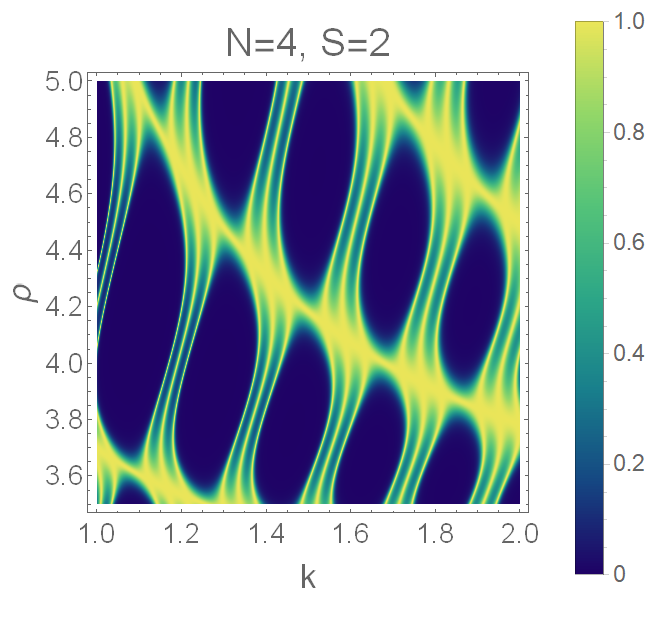} b
\includegraphics[scale=0.327]{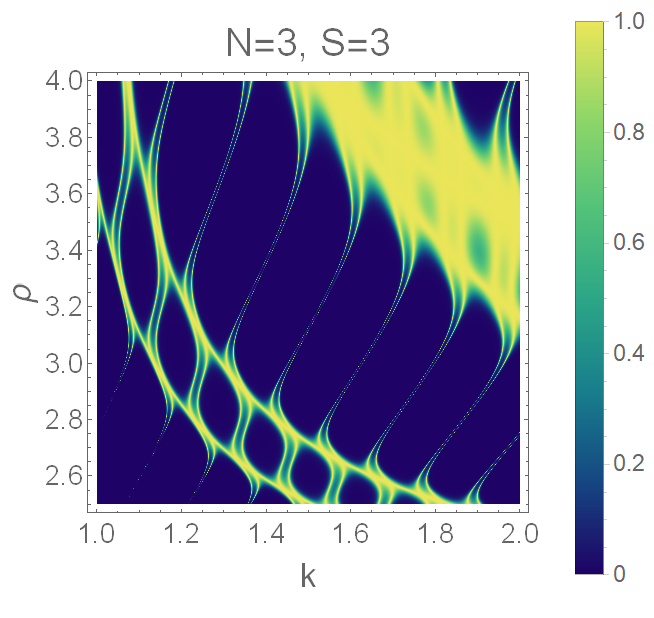} c
\includegraphics[scale=0.327]{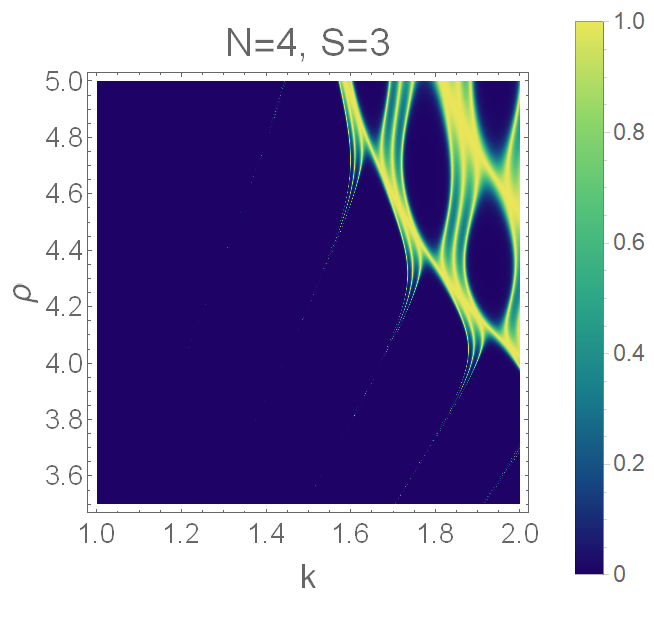} d
\caption{\it
Density plots illustrate the transmission probability in $\rho-k$ space for the CDC-$\rho_{3}$ and CDC-$\rho_{4}$ system of stage $S=2$ and $S=3$, with potential parameters set to $V=1$ and $L=50$. In all these figures transmission resonances are separated by null regions where $T(k)=0$, showing complete transmission suppression in these regions. Extremely sharp transmission resonances can be seen in Figs. (c) and (d).}
\label{figure12}
\end{center}
\end{figure}

\begin{figure}[H]
\begin{center}
\includegraphics[scale=0.315]{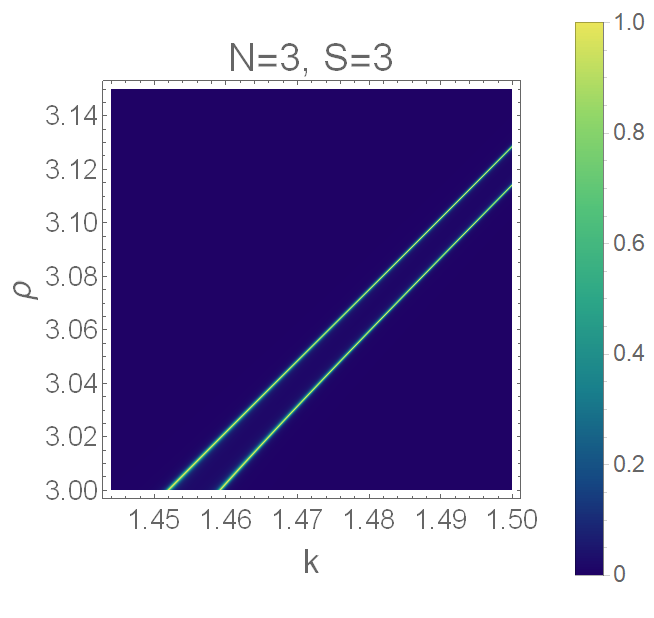} a
\includegraphics[scale=0.318]{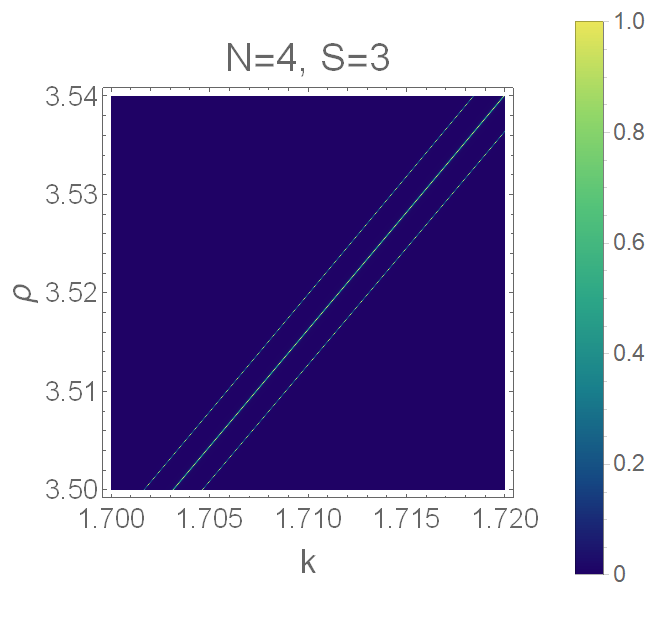} b
\caption{\it
Density plots showing (a) a zoomed-in view of a portion of Fig. $\ref{figure12}$c and (b) a zoomed-in view of a portion of Fig. $\ref{figure12}$d.}
\label{figure13}
\end{center}
\end{figure}

Fig. \ref{figure12} is crafted to show the variation in the transmission profile in the $\rho-k$ plane, through the density plots for the CDC-$\rho_{3}$ system (first column) and the CDC-$\rho_{4}$ system (second column) at stages $S=2$ (first row) and $S=3$ (second row). The potential parameters are fixed at $V=1$ and $L=50$. These density plots reveal the presence of transmission resonances interspersed with regions of near-zero transmission, $T(k) \rightarrow 0$, forming valleys in the $\rho-k$ plane. The appearance of these valleys in the transmission profile is noted as a precursor to the development of allowed and forbidden energy bands in locally periodic delta potential \cite{griffiths1992scattering}. In these density plots, the band-like feature can be seen for tor the CDC-$\rho_{3}$ and CDC-$\rho_{4}$ system of stage $S=2$ and stage $S=3$, indicating the formation of band-like structures within the Cantor-structured Dirac comb potential. Bands emerge in locally periodic delta potentials, even when the number of delta potentials is just five \cite{griffiths1992scattering, griffiths2001waves}. Similarly, in the framework of space fractional quantum mechanics (SFQM), it has been reported that such bands emerge when the number of potentials is reduced to four, with the effect being more pronounced for lower values of the L\'{e}vy index $\alpha$ \cite{tare2014transmission}. In the present study, CDC-$\rho_{2}$ system of stage $S=2$ is the simplest case of CDC-$\rho_{N}$ family, which has four ($2N^{S-1}=2\times2^{2-1}=4$) delta potential in a periodic ($\rho=\rho_{p}$) and super periodic ($\rho \neq \rho_{p}$) form. The transmission profile for this system is presented in Fig. \ref{figure08}a. Especially, this plot is intended to show the variation in the transmission profile as $\rho$ hovered around $\rho_{p}$, however, the emergence of band-like features are also observed. Similarly, band features are seen for CDC-$\rho_{3}$ system of stage $S=2$, which contains six ($2N^{S-1}=2\times3^{2-1}=6$) delta potentials. Observations from Fig. \ref{figure08} and Fig. \ref{figure12} reveal the emergence of band-like features for the Cantor-structured Dirac comb potentials, i.e. CDC-$\rho_{N}$ potential family.
On more important finding in this study is the appearance of extremely sharp transmission resonances. Fractal potentials are known for exhibiting sharp transmission resonances, a feature also observed in the present study. In Fig. \ref{figure12}, the transmission profile for stage \(S=3\) reveals sharper transmission resonances compared to stage \(S=2\), while the \(N=4\) system shows more sharp transmission than the \(N=3\) system. Exceedingly slender yellow streaks mark the locus of sharp transmission. To provide a more detailed view of these sharp yellow streaks, a magnified section of Fig. \(\ref{figure12}\)c (\(\rho\) ranging from 3.0 to 3.15, and \(k\) ranging from 1.446 to 1.50) and Fig. \(\ref{figure12}\)d (\(\rho\) ranging from 3.5 to 3.54, and \(k\) ranging from 1.7 to 1.72) is shown in Fig. \(\ref{figure13}\). These streaks manifested as abrupt transitions in the transmission probability, $T(k) = 0$ to $T(k) = 1$. Sharp transmission features from members of the Cantor potential family have been previously reported in \cite{umar2023quantum, singh2023quantum, vsingh2023quantum}. However, to the best of our knowledge, this work is the first to observe sharp transmission resonances through the Cantor-structured Dirac comb potential. This finding indicates a link between the Cantor geometry and the occurrence of very sharp transmission resonances.

\subsection{Scaling law for reflection coefficient and scaling function}
\begin{figure}[H]
\begin{center}
\includegraphics[scale=0.22]{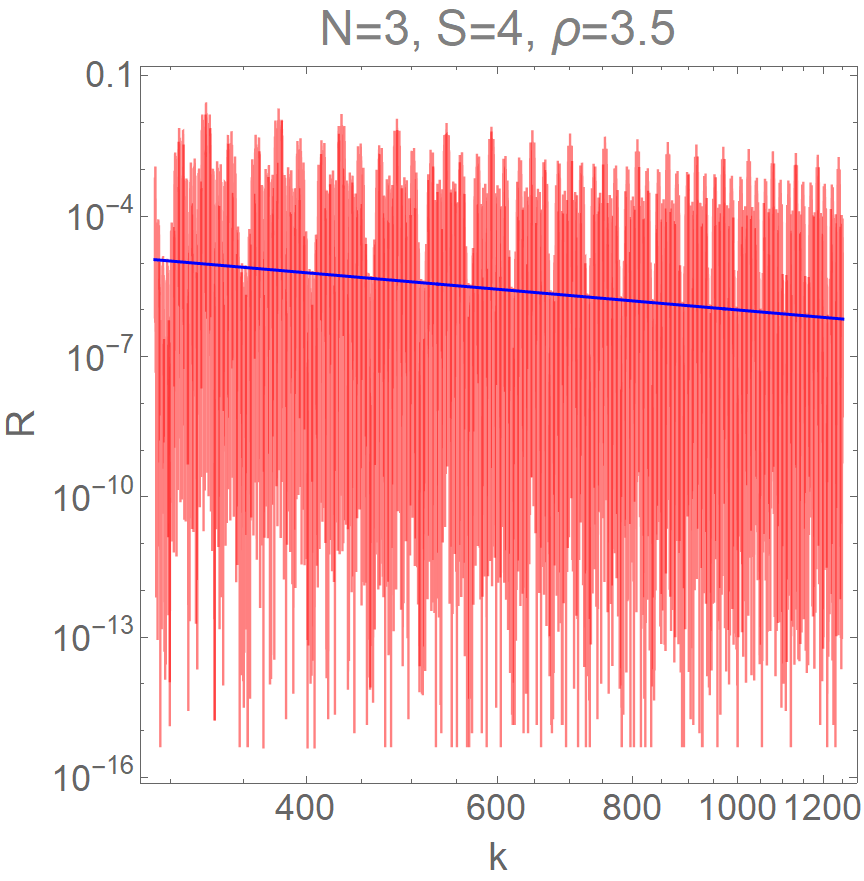} a
\includegraphics[scale=0.222]{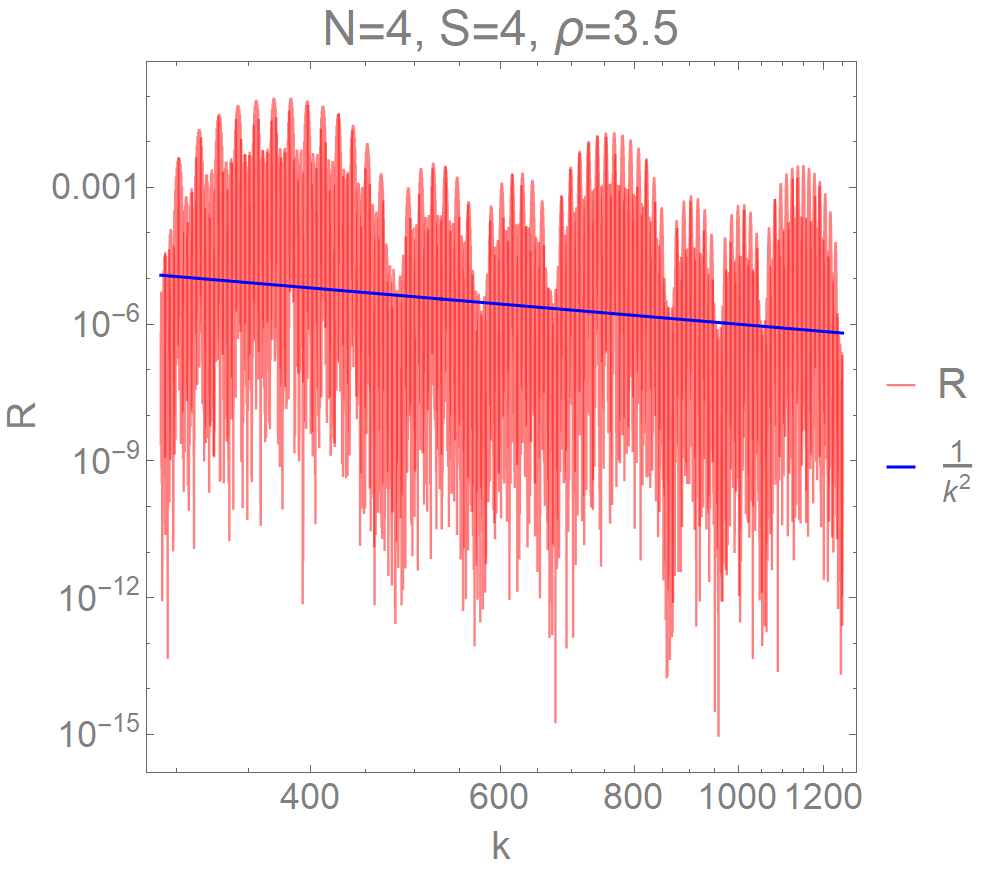} b
\caption{\it 
log-log plot representation of the scaling behavior of the reflection coefficient is depicted, which falls off according to $\frac{1}{k^2}$ (blue line). Here the potential parameters are set as $V=1$ and $L=20$.}
\label{figure14}
\end{center}
\end{figure}

The reflection coefficient for the SPP system is expressed through
\begin{equation}
    R(N_{1}, N_{2}, N_{3},\ldots,N_{S}, k)=\frac{\vert M_{12}(k)\vert^{2} \prod^{S}_{q=1}\left[U_{N_{q}-1}(\Gamma_{q})\right]^{2}}{1+\vert M_{12}(k)\vert^{2} \prod^{S}_{q=1}\left[U_{N_{q}-1}(\Gamma_{q})\right]^{2}}.
\label{scaling_02}
\end{equation}
For large $k$, the reflection coefficient is very small and can be approximated as 
\begin{equation}
    R(N_{1}, N_{2}, N_{3},\ldots,N_{S}, k) \sim \vert M_{12}(k)\vert^{2} \prod^{S}_{q=1}\left[U_{N_{q}-1}(\Gamma_{q})\right]^{2}.
\end{equation}
Therefore, for large $k$, the reflection coefficient for the super periodic delta potential is expressed through 
\begin{equation}
    R(N_{1}, N_{2}, N_{3},\ldots,N_{S}, k) \sim \frac{V^2}{k^2} \prod^{S}_{q=1}\left[U_{N_{q}-1}(\Gamma_{q})\right]^{2}=\frac{V^2}{k^2} \prod^{S}_{q=1}L^{(S)}_{q}(\gamma_{q}),
    \label{scaling_03}
\end{equation}
where $L^{(S)}_{q}(\gamma_{q})$ is the Laue function (see Eq. (\ref{eq10})) and $\gamma_{q}=\cos^{-1}\Gamma_{q}$. From the definition of CDC-$\rho_{N}$ system, it is evident that first periodic count $N_{1}=2$ and further super periodic counts $N_{2, 3, 4,\dots,S}=N$, therefore by using the property of the Chebyshev polynomial of the second kind ($U_{1}(z)=2z$) above equation can be simplified as
\begin{equation}
    R(2, N, k) \sim \frac{(2V\Gamma_{1})^2}{k^2} \prod^{S}_{q=2}\left[U_{N-1}(\Gamma_{q})\right]^{2}.
\end{equation}
It is evident from the above equation that for the CDC-\(\rho_{N}\) system, reflection coefficient \(R(2, N, k)\) exhibits a scaling behavior proportional to \(\frac{1}{k^2}\). More generally, Eq. (\ref{scaling_03}) shows that this \(\frac{1}{k^2}\) scaling is preserved in the case of the super-periodic delta potential and the scaling function
\begin{equation}
    W_{S}(k)=\prod^{S}_{q=1}\left[U_{N_{q}-1}(\Gamma_{q})\right]^{2} =\prod^{S}_{q=1}L^{(S)}_{q}(\gamma_{q}).
    \label{scaling_05}
\end{equation}
is a finite product of the Laue function. This scaling relation has previously been derived for the Cantor potential system and is documented in the literature \cite{sakaguchi2017scaling}. Moreover, this scaling behavior is also derived for other members of the Cantor potential family 
\cite{narayan2023tunneling, vsingh2023quantum, umar2023quantum, singh2023quantum, umar2024polyadic}. However, this is the first work to show that such a scaling relation also applies to the Cantor-structured Dirac comb potential and, more generally, to the super periodic delta potential. Fig. \ref{figure14}a and \ref{figure14}b present the scaling behavior of the CDC-$\rho_3$ and CDC-$\rho_4$ systems of stage $S = 4$ via a log-log plot. These plots demonstrate the behavior of the reflection coefficient for large $k$, following a scaling law of $\frac{1}{k^2}$. This scaling behavior
\begin{figure}[H]
\begin{center}
\includegraphics[scale=0.29]{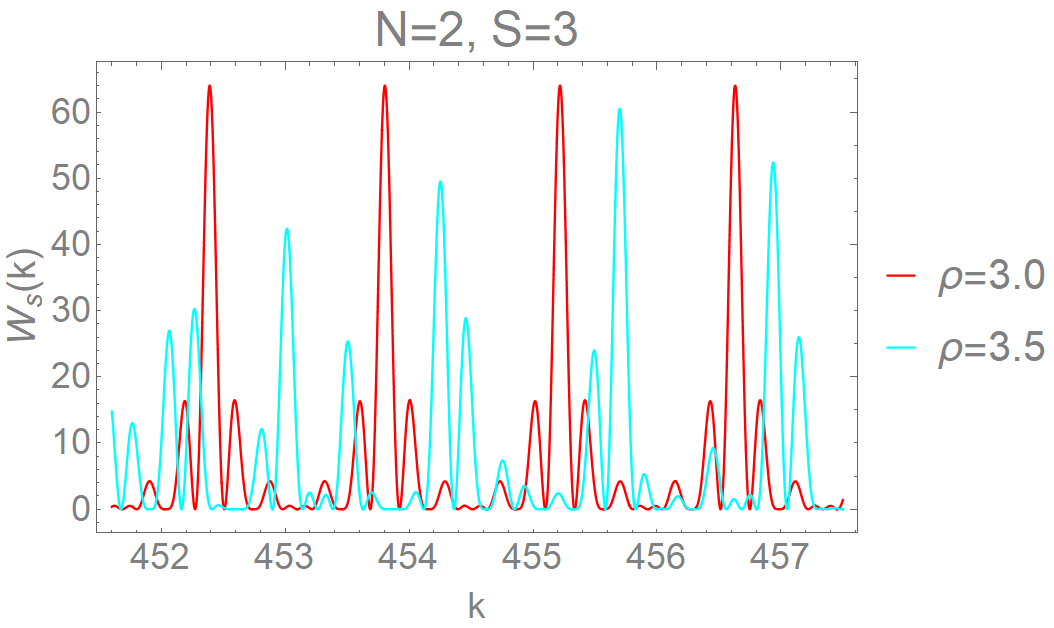} a
\includegraphics[scale=0.29]{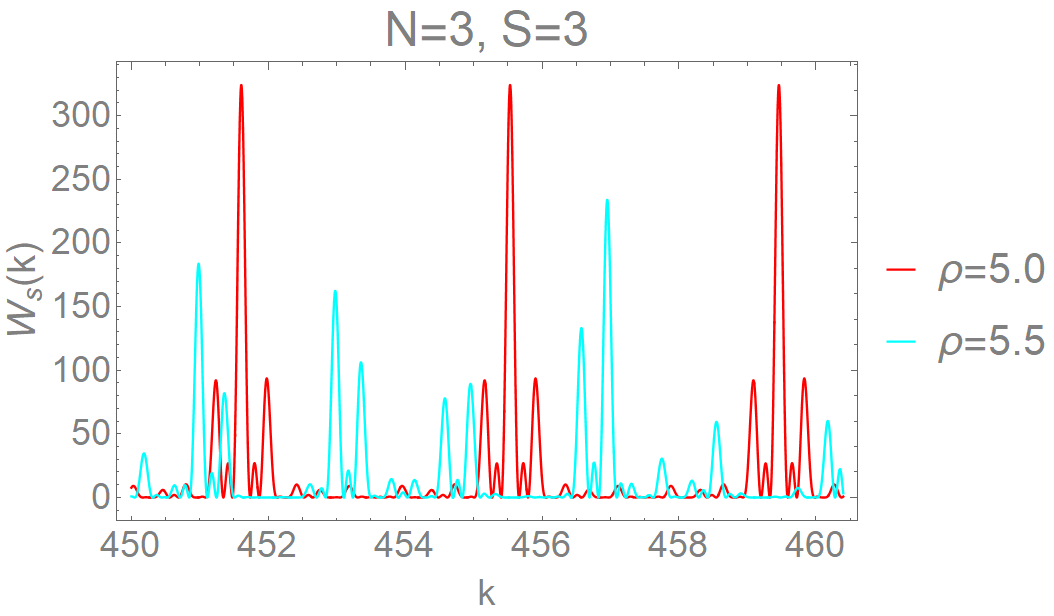} b
\caption{\it 
Variation of the scaling function $W_{3}(k)$ with $k$ for (a) $N=2$ and (b) $N=3$ for $\rho=\rho_{p}$ and $\rho=(2\rho_{p}+1)/2$, where $\rho_{p}=2N-1$. Potential parameters are $V=1$ and $L=20$.}
\label{figure15}
\end{center}
\end{figure}
\noindent
scaling behavior is governed by the scaling function $W_{S}(k)$, as described in Eq. (\ref{scaling_05}). For the CDC-$\rho_{2}$ system, the scaling function as derived from Eq. (\ref{scaling_05}) is expressed through 
\begin{equation}
    W_{3}(k) =64\Gamma_{1}^2\Gamma_{2}^2\Gamma_{3}^2.
\end{equation}
For the CDC-$\rho_{3}$ system, the corresponding scaling function is expressed as
\begin{multline}
    W_{3}(k) =4\Gamma_{1}^2\Big[64\Gamma_{2}^2\Gamma_{3}^2(1+4\Gamma_{2}^2\Gamma_{3}^2)-8(\Gamma_{2}^2+\Gamma_{3}^2)(1+16\Gamma_{2}^2\Gamma_{3}^2)+16(\Gamma_{2}^4+\Gamma_{3}^4)+1 \Big].
\end{multline}
Fig. \ref{figure15} illustrates the variation of the scaling function $W_3(k)$ for both the CDC-$\rho_2$ and CDC-$\rho_3$ systems, with \(\rho = \rho_p\) and $\rho = (2\rho_p + 1)/2$. Here, \(\rho_p = 2N - 1\), which determines the periodicity of the delta potentials at the second stage of the CDC-\(\rho_N\) system. When \(\rho = \rho_p\), periodicity is observed in the scaling function and this periodicity is lost when \(\rho \neq \rho_p\). For large \(k\), the reflection coefficient is directly influenced by this scaling function. A deeper investigation of this scaling function is required, which presents a promising avenue for further research.

\section{Conclusion}
\label{section08}
In the present study, we introduce the Cantor structured Dirac comb potential concept, a novel super periodic delta potential, referred to as the Cantor Dirac comb (CDC-$\rho_{N}$) potential system. The CDC-$\rho_{N}$ system is constructed by positioning the delta potentials at the boundaries of the rectangular potential segments at each stage of the fractal Cantor/polyadic Cantor potential. In this model, the Dirac comb system of any stage $S$ is confined in the interval [$0, L$]. We demonstrate that, akin to the fractal Cantor potential and the low lacunarity polyadic Cantor potential, the CDC-$\rho_{N}$ system represents a specific case of the SPP system. Utilizing the SPP formalism, we derive the expression for the transmission coefficient for this system. Since the CDC-$\rho_{N}$ system is a special case of the SPP system, we explore the characteristics of the SPP system, such as transmission resonances and the formation of resonance bands. Subsequently, we discuss the transmission through the CDC-$\rho_{N}$ system.\\
\indent
This novel potential system allows us to show the behavior of the transmission profile as the delta potential system transits from a periodic to a super periodic system. A noteworthy observation is the appearance of band-like features, which may serve as precursors to the formation of energy bands within the Cantor-structured Dirac comb potential. Moreover, a critical and noteworthy aspect of our investigation is the identification of exceptionally sharp transmission resonances within this newly introduced potential system which are usually not seen in the other Hermitian systems except the fractal potentials. This finding suggests a strong connection between the fractal Cantor geometry and the appearance of sharp transmission. The occurrence of such sharp resonances suggests promising applications in the development of transmission filters with ultra-narrow bandwidth, offering precise wavelength selectivity.\\
\indent
In our exploration, we have studied the reflection coefficient \( R(k) \) for large values of \( k \), revealing a result where \( R(k) \) scales as \( \frac{1}{k^{2}} \). This scaling behavior has been validated through both analytical derivations and graphical representations. Generally, this scaling property is adhered to by the super-periodic delta potential, and consequently, by the CDC-\(\rho_{N}\) system. Notably, this behavior is governed by the scaling function \( W_{S}(k) \), which is expressed as a finite product ($S$ times) of the Laue function. An intriguing aspect of this scaling function is its periodicity with respect to $k$ when the CDC-\(\rho_{N}\) system is such that the delta potential is periodic (\( \rho = \rho_{p} \)) at the second stage, whereas the periodicity is lost as the delta potential transitions to a super-periodic system (\( \rho \neq \rho_{p} \)) at this stage. Future work could offer deeper insights by further exploring the characteristics of this scaling function, with a particular focus on investigating its fractal nature.\\
\indent
The presented CDC-$\rho_{N}$ system exemplifies a fractal-structured Dirac comb potential, as it construction lies upon both Cantor potentials ($N=2$) and low lacunarity polyadic Cantor potentials ($N>2$), which are categorized as symmetric fractals. The exploration of non-fractal structured Dirac comb potentials and asymmetric fractal-structured Dirac comb potentials represents a promising direction for future research. Recently, we introduced the concept of the generalized \textit{unified} Cantor potential (UCP-$\rho_{N}$) system \cite{umar2024polyadic}, which bridges the gap between fractal and non-fractal potentials. This framework provides a foundation for future research on the transmission of quantum waves through the \textit{unified} Cantor-structured Dirac comb potential. In particular, it offers a unique opportunity to explore how the transmission characteristics evolve as the system transitions from a fractal to a non-fractal structure, offering deeper insights into the behavior of such potentials. Furthermore, the CDC-$\rho_{N}$ system can be examined within the context of space fractional quantum mechanics (SFQM). Additionally, the Cantor-structured non-Hermitian Dirac comb potential presents an intriguing avenue for study within non-Hermitian quantum mechanics (NHQM), where novel physical phenomena related to wave propagation and quantum transport in complex media may emerge. In summary, research on the CDC-$\rho_{N}$ system, with its potential extensions into both fractal and non-fractal realms, as well as its application in SFQM and NHQM, opens new avenues for exploration. Such studies are essential for advancing our understanding of quantum mechanics and could yield both theoretical insights and practical innovations\\
\\
\\
\\
\\
\noindent
{\it \bf{Acknowledgements}}:\\
\\
MU gratefully acknowledges Prof. P. Senthilkumaran, Head, Optics and Photonics Centre (OPC), Indian Institute of Technology (IIT) Delhi, for his support and guidance. MU also expresses deep appreciation to the OPC at IIT Delhi for its essential role in creating a supportive research environment.

\newpage
\bibliographystyle{elsarticle-num}
\bibliography{References}

\end{document}